\documentclass[12pt,preprint]{emulateapj}

\usepackage{multirow,color,natbib,wrapfig,ulem,graphicx}

\newcommand{\LCDM}{$\Lambda$CDM}

\newcommand{\vmaxdmo}{$v_{max,dmo}$}

\newcommand{\twovmax}{$2\,v_{max,dmo}$}
\newcommand{\twovmaxsini}{$2\,v_{max,dmo}\,\sin i$}
\newcommand{\wfifty}{$w_{50}$}
\newcommand{\wfiftye}{$w_{50}^e$}

\newcommand{\kms}{km$\,$s$^{-1}$}

\newcommand{\Msun}{\mbox{\,$M_{\odot}$}}

\def\spose#1{\hbox to 0pt{#1\hss}}
\def\simlt{\mathrel{\spose{\lower 3pt\hbox{$\mathchar"218$}}
     \raise 2.0pt\hbox{$\mathchar"13C$}}}
\def\simgt{\mathrel{\spose{\lower 3pt\hbox{$\mathchar"218$}}
     \raise 2.0pt\hbox{$\mathchar"13E$}}}

\slugcomment{Accepted for publication in ApJ}

\shorttitle{Interpreting the Galaxy VF}
\shortauthors{Brooks et al.}

\begin{document}


\title{How to Reconcile the Observed Velocity Function of Galaxies with Theory}


\author{Alyson\ M.\ Brooks\altaffilmark{1}, Emmanouil\ Papastergis\altaffilmark{2,$\dag$}, Charlotte R. Christensen\altaffilmark{3}, Fabio Governato\altaffilmark{4}, Adrienne Stilp\altaffilmark{5}, Thomas R. Quinn\altaffilmark{4}, James Wadsley\altaffilmark{6}}


\altaffiltext{1}{Department of Physics \& Astronomy, Rutgers University, 136 Frelinghuysen Rd., Piscataway, NJ 08854; 
  \email{abrooks@physics.rutgers.edu}}
\altaffiltext{2}{Kapteyn Astronomical Institute, University of Groningen, Landleven 12, Groningen NL-9747AD, Netherlands;
  \email{papastergis@astro.rug.nl}}
\altaffiltext{$\dag$}{\textit{NOVA} postdoctoral fellow}
\altaffiltext{3}{Department of Physics, Grinnell College, Noyce Science Center, 1116 Eighth Ave., Grinnell, IA 50112}
\altaffiltext{4}{Department of Astronomy, Box 351580, University of Washington, Seattle, WA 98195-1580}
\altaffiltext{5}{Department of Biostatistics, Box 359461, University of Washington, 4333 Brooklyn Ave. NE, Seattle, WA 98195-9461}
\altaffiltext{6}{Department of Physics \& Astronomy, McMaster University, Hamilton, ON, L8S 4M1, Canada}

\date{\today}


\begin{abstract}
Within a $\Lambda$ Cold Dark Matter ($\Lambda$CDM) scenario, we use high resolution cosmological simulations spanning over four orders of magnitude in galaxy mass to understand the deficit of dwarf galaxies in observed velocity functions.  We measure velocities in as similar a way as possible to observations, including generating mock HI data cubes for our simulated galaxies.  We demonstrate that this apples-to-apples comparison yields an ``observed'' velocity function in agreement with observations, reconciling the large number of low-mass halos expected in a $\Lambda$CDM cosmological model with the low number of observed dwarfs at a given velocity.  We then explore the source of the discrepancy between observations and theory, and conclude that the dearth of observed dwarf galaxies is primarily explained by two effects. The first effect is that galactic rotational velocities derived from the HI linewidth severely underestimate the maximum halo velocity. 
The second effect is that a large fraction of halos at the lowest masses are too faint to be detected by current galaxy surveys. We find that cored dark matter density profiles can contribute to the lower observed velocity of galaxies, but only for galaxies in which the velocity is measured interior to the size of the core ($\sim$3 kpc). 
\end{abstract}



\section{Introduction}

\setcounter{footnote}{0}

The velocity function of galaxies is indicative of the number of galactic halos that exist as a function of mass, and is therefore a powerful test of our cosmological galaxy formation model. For galaxies with velocities above $\sim$100 \kms ~that are primarily dispersion-dominated, the observed velocity function (VF) is generally in agreement with theoretical expectations within a $\Lambda$ Cold Dark Matter (CDM) cosmology, as long as the effects of baryons are included \citep{Gonzalez2000, Sheth2003, Chae2010, Obreschkow2013}.  To probe to lower galaxy masses which are more likely to be rotation-dominated, HI rotation data is ideal.  
\citet{Zwaan2010} and \citet{Trujillo-Gomez2011} were some of the first to combine the early type galaxy VF from SDSS with the HIVF from HIPASS to probe to lower galaxy masses, and found that theory predicted more dwarfs below $\sim$80 \kms \ than observed \citep[see also][]{Abramson2014, Bekeraite2016}.

The HIVF has since been updated, thanks in large part to data from the ALFALFA HI survey \citep{Giovanelli2005}, and from systematic optical searches for neighboring galaxies \citep{Karachentsev2013}.
\citet{Papastergis2011} used early data from the ALFALFA survey (at 40\% of its eventual sample size) to confirm that there is a deficit of low mass observed galaxies compared to that expected in a $\Lambda$CDM cosmology. 
In galaxies with rotational velocities, $v_{rot}$, $\sim$25 \kms, the ALFALFA HIVF shows nearly an order of magnitude fewer galaxies than expected based on straightforward $\Lambda$CDM estimates (e.g., that each dark matter halo contains one luminous galaxy).  
\citet{Klypin2015} made a separate measurement of the VF using the catalog of Local Volume galaxies out to 10 Mpc \citep{Karachentsev2013}.  They derived velocities for galaxies as faint as $M_B = -10$. 
Despite probing to these low masses and correcting for completeness, they still found a dearth of low velocity galaxies compared to the number expected in CDM. Likely due to the fact that they could include gas-poor faint galaxies, the discrepancy is not as large as seen in the ALFALFA HIVF sample, but they confirm the nearly factor of 10 discrepancy between theory and observation at $v_{rot} \sim25$ \kms.

This missing dwarf problem is reminiscent of the missing satellites problem \citep{Moore1999, Klypin1999}, but now extends into the field, beyond the virial radius of more massive galaxies. This means that solutions that rely on the tidal field of the host galaxy to reduce the numbers and masses of satellite dwarfs \citep{Zolotov2012, Brooks2013, Arraki2014, BZ2014, Wetzel2016} should not apply, and a new mechanism to reduce the number of field galaxies needs to be invoked.  One long-standing solution to the missing dwarf problem is warm dark matter \citep[WDM, e.g.,][]{Bode2001, Polisensky2011, Menci2012, Lovell2012, Nierenberg2013}, in which the thermal relic mass of the dark matter particle is $\gtrsim 2$ keV. 
However, \citet{Klypin2015} and \citet{Brook2015} showed that WDM more massive than 1.5 keV doesn't suppress enough structure at low masses to be compatible with the observed VF \citep[though see][]{Schneider2016}.  Lighter masses have already been ruled out based on the small scale structure observed in the Lyman-$\alpha$ forest at high redshift \citep{Viel2006, Seljak2006, Viel2008, Viel2013}. Hence, WDM is difficult to make compatible with all available observational constraints.

Another interpretation of the observed VF is not that there are dwarfs missing, but that low mass galaxies display lower velocities than anticipated.  This may be due to complications related to the way rotational velocities are measured observationally \citep{Brook2016, Maccio2016, Yaryura2016}, baryonic physics \citep[e.g.,][]{Brook2014,Brook2015}, or to dark matter physics if the dark matter is self-interacting \citep{Spergel2000, Loeb2011, Vogelsberger2012, Zavala2013, Elbert2014, Fry2015}.  
However, assigning galaxies with low HI rotational velocities to relatively large halos in order to reproduce the observed VF has also proven challenging in the $\Lambda$CDM context. This is because the internal kinematics of dwarfs seem to indicate low-mass hosts. \citet{Ferrero2012} demonstrated that galaxies with stellar masses in the 10$^6$-10$^8$ M$_{star}$ range appear to be hosted by much smaller halos than predicted by abundance matching, based on their observed rotation velocities.  \citet{Papastergis2015} extended this to a much larger sample, but confirmed that galaxies with HI rotation velocities below $\sim$25 \kms \ were incompatible with residing in the more massive halos that abundance matching predicts.  \citet{Garrison-Kimmel2014} used the observed densities of local group dwarf irregulars to derive the halo masses that they reside in.  
All of the galaxies seemed to be in halos of similar mass, %\citep[see also][]{Ferrero2012}, 
but it was a much lower mass than predicted by abundance matching.  They concluded that it does not seem possible to simultaneously reproduce the measured velocity function (i.e., satisfy abundance matching) and the observed densities of galaxies, an issue referred to as the too big to fail problem. 

However, recent work by \citet{Brook2014,Brook2015} used results from simulations in which stellar feedback processes alter the dark matter content of dwarf galaxies to show that they can simultaneously match the densities and velocities of observed dwarfs.  In this scenario, feedback from stars and supernovae create bursty star formation histories in dwarf galaxies that fluctuate the gravitational potential well at the center of the dwarf \citep{Pontzen2012, Teyssier2013, Pontzen2014, Onorbe2015, Chan2015, Dutton2016}.  Dark matter core creation leads to a better match between theory and observed rotation curves \citep{Katz2017, Santos2017}.  Feedback is particularly effective in dwarf galaxies with halo masses of a few 10$^{10}$ M$_{\odot}$ \citep{Governato2012, diCintio2014a}, where it can transform an initially steep inner dark matter density profile into a flatter ``cored'' profile.  At lower halo masses there is less star formation, leading to less energy injection and lower core formation efficiencies \citep{Penarrubia2012, Maxwell2015}.  At higher masses, the deeper potential wells of galaxies make core formation increasingly difficult \citep{diCintio2014a, Pontzen2014}, at least if an additional source of feedback is neglected, such as AGN \citep{Martizzi2013}.  In this model with baryonic feedback, it is possible to assign dwarf galaxies to relatively massive halos, despite the low rotational velocities measured from their spatially resolved stellar kinematics. This is because baryonic feedback can push dark matter out of the central regions, lowering the enclosed mass at the radii that stellar kinematics probe (but without affecting the total halo mass\footnote{modulo a slight reduction in halo mass caused by the loss of baryons or preventive feedback \citep{Munshi2013, Sawala2013}}).  Hence, the densities are lowered, and the apparent velocities of the galaxies, reconciling the observations with theory.

Based on such simulations, \citet{diCintio2014b} derived an analytic model for the dark matter density profile that varies with stellar-to-halo mass ratio.  \citet{Brook2014} and \citet{Brook2015} used this analytic model to derive galaxy trends that they claim reconcile the halo densities and the observed VF.  In this work, we use simulations directly.  These simulations also create dark matter cores \citep{Governato2012, Zolotov2012, Christensen2014a, BZ2014}, following very similar trends to those in \citet{diCintio2014b}.  However, because we use the simulations directly, we do not have to resort to analytic models for the baryon distribution in the galaxies.  \citet{Maccio2016} also recently used simulations directly to show that baryonic simulations can be reconciled with observations.  However, they did not investigate the role of dark matter cores in their results. We show that accounting for the gas distribution is important and not straightforward.  Unlike \citet{Brook2015}, we do not find that dark matter core creation consistently has a large impact on observed velocities of galaxies, yet we {\it do} find that we can reproduce the observed VF.  

This paper is organized as follows: In Section 2 we present information about the simulations.  In Section 3 we demonstrate that deriving velocities from baryons yields a substantially lower velocity in dwarf galaxies than expected from theoretical results that rely on dark matter-only simulations.  In Section 3.1 we explore how completeness (i.e., the number of detectable halos at low velocity) affects the observed VF. In Section 3.2 we describe our method to mimic observations and derive velocities in as close a way as possible to the observations. In Section 3.3 we re-derive the expected VF given our completeness results and mock observed velocities.  In Section 3.4 we demonstrate that our simulations match other essential scaling relations.  We systematically explore the importance of various effects in reducing the observed velocities relative to to theoretical velocities in Section 4. In Section 5 we explore the role of dark matter cores on the reduced observed velocities.  We find that cores are only important in galaxies where the velocity is measured interior to the size of the core.  We compare are results to previous work in Section 6, and conclude in Section 7.

\section{The Simulations}
\label{sims}

\begin{deluxetable*}{ccccccccc}
%\centering
% \begin{minipage}{140mm}
%  \title{Simulated Galaxy Properties \label{obssum}}
\tabletypesize{\footnotesize}
\tablecaption{Properties of the Simulated Galaxies}
\tablehead{
\colhead{Simulation}  &  
\colhead{$v_{max,dmo}$ range}   &  
\colhead{M$_{star}$ range}  &  
\colhead{m$_{DM,part}$}    &
\colhead{m$_{star,part}$}  &  
\colhead{Softening}   &  
\colhead{Overdensity} &  
\colhead{N$_\textrm{DM}$}   	      \\
 & \kms & \Msun & \Msun & \Msun   &  pc & $\Delta \rho /\rho$ &  within R$_{vir}$  \\
 (1) & (2) & (3) & (4) & (5) & (6) & (7) & (8) } 
\startdata
Fields 1 -- 6   & 30-150 & 2$\times$10$^5$-10$^{10}$       & 1.6$\times$10$^{5}$  & 8 $\times $10$^{3}$ & 174 & -0.15 to 1.35 & 0.03-3.4$\times$10$^{6}$  \\
Field 7         & 43-56  & 2$\times$10$^7$-3$\times$10$^8$ & 2$\times$10$^{4}$   & 10$^{3}$             & 85 & -0.02 & 0.05-2$\times$10$^{6}$   \\
Field 8         & 38     & 10$^8$                          & 6$\times$10$^{3}$   & 4.2$\times$10$^{2}$  & 64 &  0.01 & 2$\times$10$^{6}$ 
\enddata
%\end{minipage}
\tablecomments{All fields have been run both with baryons and as DM-only. Column (2) lists the $v_{max,dmo}$ range of each galaxy at $z$=0 in the DM-only version of the run. Column (3) lists the stellar mass range of the galaxies at $z$=0 in the baryonic version of the run. Columns (4) and (5) list the mass  of individual dark matter and star particles, respectively, in the baryonic runs.   Column (6) shows  $\epsilon$, the spline gravitational force softening, in pc.  Column (7) shows the environmental density relative to the average, the {\it rms} mass fluctuation on 8h$^{-1}$ Mpc scales. 
Column (8) lists the range in total number of DM within the virial radius of the halos at $z$=0 in the baryonic runs. }
\label{table1}
\end{deluxetable*}

The high-resolution simulations used in this work were run with {\sc pkdgrav} \citep{Stadel2001} and its baryonic (SPH) version {\sc gasoline} \citep{Wadsley2004}, using a $\Lambda$ Cold Dark Matter ($\Lambda$CDM) cosmology with $\Omega_m$ = 0.24, $\Omega_{\Lambda}$ = 0.76, $H_0$ = 73 km s$^{-1}$, $\sigma_8$=0.77, and n=0.96.  The galaxies were originally selected from two uniform dark matter-only simulations of 25 and 50 comoving Mpc per side.  From these volumes, eight field--like regions where selected, each centered on a galaxy with halo mass\footnote{The virial radius is defined relative to critical density, $\rho_{c}$, where the mean density enclosed is $\rho/\rho_{c} \approx 100$ at $z=0$.} ranging from 10$^{10}$ to 10$^{12}$ M$_{\odot}$.  Each field was then resimulated using the ``zoom-in'' volume renormalization technique \citep{Katz1993}, which simulates a region out to roughly 1 Mpc of the primary halo at the highest resolution, while fully preserving the surrounding large scale structure that builds angular momentum in tidal torque theory \citep{Peebles1969,Barnes1987}. These simulations were run from approximately $z=150$ to $z=0$. A uniform UV background turns on at $z = 9$, mimicking cosmic reionization following a modified version of \citet{Haardt2001}.  The {\it rms} mass fluctuation relative to the cosmic average, $\delta \rho$/$\rho$, for each chosen field ranges from -0.15 to1.35 when measured on a scale of 8h$^{-1}$ Mpc (see Table~\ref{table1}).  Five of the fields fall within 0.05 standard deviations of the cosmic mean density. 

The spline force softening, $\epsilon$, ranges from 64 pc to 174 pc in the high resolution regions (see Table~\ref{table1}), and is kept fixed in physical pc at $z < 10$.  The dark matter (DM) and stellar mass resolutions are listed in Table~\ref{table1}.  The gas smoothing length is allowed to shrink as small as 0.1$\epsilon$ in very dense regions (0.5$\epsilon$ is typical) to ensure that hydro forces dominate at very small scales. The main galaxy in every zoomed region contains several millions of DM particles within its virial radius.  

The high resolution of these cosmological simulations allows us to identify the high density peaks where H$_2$ can form.  We track the non-equilibrium formation and destruction of H$_2$, following both a gas-phase and a dust (and hence metallicity) dependent scheme that traces the Lyman-Werner radiation field and allows for gas and dust self-shielding \citep{Gnedin2009, Christensen2012}.  We include cooling from both metal lines and H$_2$ \citep{Shen2010, Christensen2012}.  Metal cooling, H$_2$ fractions, and self-shielding of high density gas from local radiation play an important role in determining the structure of the interstellar medium and where star formation can occur \citep{Kennicutt1998, Krumholz2008,Bigiel2008, Blanc2009, Bigiel2010, Schruba2011, Gnedin2011,Narayanan2012}. With this approach, we link the {\it local} star formation efficiency directly to the local H$_2$ abundance.  As described in \citet{Christensen2012}, the efficiency of star formation, $c^*$, is tied to the H$_2$ fraction, $X_{\rm H_2}$.  The resulting star formation rate (SFR) depends on the local gas density such that SFR $\propto$ $c^*$X$_{\rm H_2}(\rho_{\rm gas})^{1.5}$, with $c^* = 0.1$.  This value of $c^*$ gives the correct normalization of the Kennicutt-Schmidt relation.\footnote{Note that the efficiency of star formation in any given region is actually much lower than the implied 10\%, due to the fact that feedback from newly formed stars quickly disrupts gas, shuts off cooling, and lowers the overall efficiency \citep[e.g.,][]{Agertz2013}.} Because star formation is restricted to occurring in the presence of H$_2$, stars naturally form in high density regions ($>$ 100 amu cm$^{-3}$), with no star formation density threshold imposed. 

Star particles represent a simple stellar population born with a \citet{Kroupa1993} initial mass function.  The star particles lose mass through stellar winds and supernovae (SN\,Ia and SN\,II).  Supernovae deposit 10$^{51}$ ergs of thermal energy into the 
surrounding gas following the ``blastwave'' scheme described in \citet{Stinson2006}.  No velocity ``kicks'' are given to the surrounding gas particles, but thermal energy is deposited and cooling is turned off within a ``blastwave'' radius and adiabatic expansion phase calculated following \citet{Ostriker1988}.  The thermal energy deposition from supernovae can lead to bubbles of hot gas that expand, driving winds from the galaxies.  Unlike other ``sub-grid'' schemes, the gas stays hydrodynamically coupled while in galactic outflows.  Despite its reliance on supernovae, this model should be interpreted as a scheme to model the effect of energy deposited in the local interstellar medium by {\it all} processes related to young stars, including UV radiation from massive stars \citep{Hopkins2011,Wise2012,Agertz2014}.  The rate of ejected mass in winds in these simulations is dependent on galaxy mass, ranging from less than the current SFR in Milky Way-mass galaxies, to typically a few times the current SFR in galaxies with $v_{circ} \sim 50$ \kms, to more than 10 times the current SFR in galaxies with $v_{circ} \sim 20$ \kms \citep{Christensen2016}. These ejection rates are similar to what is observed in real galaxies over a range of redshifts \citep{Martin1998,Kirby2011,Kornei2012}. Additionally, \citet{Munshi2013} demonstrated that these simulations match the observed stellar mass to halo mass relation \citep{Moster2013}, by creating a more realistic star formation efficiency as a function of galaxy mass.

The star formation and feedback in these simulations leads to important trends in the resulting galaxies that are important for the present study. First, feedback strongly suppresses star formation, but the amount of suppression scales with galaxy mass \citep{Brooks2007}.  In the deeper potential wells of massive galaxies, high densities make it easier for the gas to cool quickly after being heated by supernovae.  The lower densities in dwarf galaxies are more susceptible to heating, driving the star formation efficiencies even lower in dwarfs.  Hence, even though the simulated dwarf galaxies may lose much of their gas in winds \citep{Christensen2016}, the gas that stays behind is very inefficient at forming stars, so that the dwarfs are very gas rich \citep{Munshi2013}. Second, when star formation is tied directly to high density regions with H$_2$, subsequent feedback causes these cold, dense regions to become massively over-pressurized.  This leads to very bursty star formation histories in dwarf galaxies \citep{Dominguez2014, vanderwel2011, Kauffmann2014}.  Bursty star formation creates fluctuations in the galaxy potential well, particularly in halos with masses a few 10$^{10}$ M$_{\odot}$ \citep{Governato2012, diCintio2013, diCintio2014a}, which causes initially cuspy dark matter density profiles to transform into flatter ``cores''. In Section \ref{corecreation} we examine whether this core formation lowers the measured $v_{rot}$ of halos from that predicted in DM-only simulations.

\section{The Impact of Baryons on the VF}

The goal of this study is to identify whether baryonic processes can reconcile the VF expected theoretically in a \LCDM \ universe with observations. The first attempts to compare the theoretical and observational VFs were based on the results of DM-only cosmological simulations \citep[e.g.,][]{Zavala2009,Zwaan2010,Papastergis2011,Trujillo-Gomez2011}. However, this simple approach neglects several important baryonic effects that are important for making a fair comparison between theory and observations.
Here, we analyze the impact of baryonic effects on the theoretical VF by using simulations run both with baryons and as DM-only. Our aim is to perform ``mock observations'' of our baryonic simulations, and derive a theoretical VF in as similar a way as possible to current observational determinations. 

In what follows, various definitions of velocity arise.  In order to compare results from observations to results from theory, one must define a characteristic galaxy velocity that can be compared.  In practice, each defined characteristic velocity is slightly different, being derived in a slightly different way.  Below, we explore in detail the results of various definitions of characteristic velocity.  To minimize confusion for the reader, in Table~\ref{veldef} we define each velocity that we use in the remainder of this paper.

We focus our comparison on the VF measured in the Local Volume ($D \lesssim 10$ Mpc) by \citet{Klypin2015}, based on the catalog of nearby galaxies of \citet{Karachentsev2013}. The catalog is optically selected, and probes with reasonable completeness galaxies as faint as $M_B = -10$. The majority ($\sim$80\%) of galaxies in this Local Volume catalog have measurements of their rotational velocity based on the width of their HI profile, \wfifty. Some fraction of galaxies lack HI data, either because they are intrinsically gas-poor (e.g., satellites of nearby massive galaxies), or because they have not been targeted by HI observations. These galaxies are assigned rotational velocities based on stellar kinematic measurements when available, or otherwise according to an empirical luminosity-velocity relation. Note that the \citet{Klypin2015} VF is consistent with other independent observational measurements of the VF, such as the one performed by the ALFALFA blind HI survey \citep{Papastergis2011,Papastergis2016}.

In order to make an appropriate comparison with the observational VF measured by \citet{Klypin2015}, we need to model two key observational effects using our baryonic simulations. First, we need to replicate the completeness limitations of the \citet{Karachentsev2013} catalog at low luminosities. Faint galaxies tend to have low rotational velocities. Hence, if a halo is not detectable in current surveys, the density of observed galaxies at the low velocity end of the VF will be surpressed relative to theoretical expectations that populate each halo with a detectable galaxy. Thus completeness can significantly impact the measurement of the low-velocity end of the VF and must be accounted for. Second, we need to compute the theoretical VF in terms of the rotational velocity measured observationally, \wfifty. This entails deriving realistic estimates of the HI linewidths for our baryonic halos.

\begin{deluxetable*}{ll}
\tabletypesize{\footnotesize}
\tablecaption{Characteristic Velocity Definitions in the Text}
\tablehead{ 
\colhead{symbol}  &  
\colhead{definition}  } 
\startdata
$v_{circ}$ & circular velocity; $v_{circ} = \sqrt{GM/r}$ where $M$ is the mass enclosed within radius $r$ \\ [2mm]
\vmaxdmo  & the maximum value of $v_{circ}$ for a dark matter-only simulated halo; 2\vmaxdmo ~is the theoretical counterpart to \wfiftye  \\ [2mm]
\twovmaxsini  & twice the maximum value of $v_{circ}$ measured for a dark matter-only simulated halos multiplied by the $sin$ of the \\
 & observational inclination angle $i$; the theoretical counterpart to \wfifty  \\ [2mm]
 \wfiftye  &  for galaxies with measurable HI: the full width of a galaxy's HI line profile, measured at 50\% of the profile peak \\ 
 & height when the galaxy is viewed edge-on (inclination $i$ = 90$^\circ$); for galaxies with no measurable HI: twice the \\ 
 & stellar velocity dispersion; the observational counterpart to 2\vmaxdmo  \\ [2mm]
\wfifty   & \wfiftye $\times$ sin($i$); \wfifty ~for a galaxy viewed at a random inclination angle $i$; observational counterpart to \twovmaxsini    \\ [2mm]
$w^e_{20}$ & similar to \wfiftye ~but measured at 20\% of the HI profile peak height \\ [2mm]
$V_f$ & velocity of a galaxy measured on the flat part of the rotation curve \\ [2mm]
$v_{max,sph}$ & the maximum value of $v_{circ}$ for a galaxy halo in a baryonic simulation  \\ [2mm]
$v_{out}$ & $v_{circ}$ measured at R$_{out}$, the radius at which a galaxy's HI surface density falls below 1 M$_{\odot}$/pc$^2$  \\ [2mm]
$v_{out,dmo+b}$ & $v_{circ}$ for a dark matter-only halo (reduced by a velocity consistent with removing the cosmic baryon fraction) + $v_{circ}$ \\
 & for only the baryons in the counterpart baryonic simulation. Measured at R$_{out}$, where R$_{out}$ is determined from the \\
 & simulated baryonic counterpart
\enddata
%\end{minipage}
\tablecomments{Note that \wfifty, \wfiftye, and $w_{20}$ are all derived from spatially unresolved data.  The remainder of the characteristic velocities are derived from spatially resolved data, and are associated with a particular radius within a given galaxy.} 
\label{veldef}
\end{deluxetable*}

\subsection{Detectability of halos}
\label{detectability}

Each ``zoomed'' simulation contains a high resolution region centered on a halo, ranging in virial mass from 10$^{10}$ M$_{\odot}$ to 10$^{12}$ M$_{\odot}$. In addition to the central halo, every zoomed region contains smaller galaxies that we also include in our analysis. Because the theoretical VF is traditionally derived using results from DM-only simulations, for every simulated baryonic halo we identify its counterpart in the DM-only run in order to assign a \vmaxdmo \ value, the maximum circular velocity in the DM-only runs. Because the DM particles are identical in both the baryonic and DM-only initial conditions, identifying a counterpart is relatively straightforward. For all halos in the DM-only run with more than 64 particles, we identify the DM particles that make up each halo in the DM-only run, then find those same particles in the baryonic run and note the halo\footnote{Halos are identified with AHF, {\bf{A}}MIGA's {\bf{H}}alo {\bf{F}}inder \citep{Gill2004, Knollmann2009}. AHF is available for download at \texttt{http://popia.ft.uam.es/AMIGA/}.} that most of those particles belong to.  We find a matching counterpart for 6271 halos and subhalos.

\begin{figure}
\includegraphics[angle=0,width=\columnwidth]{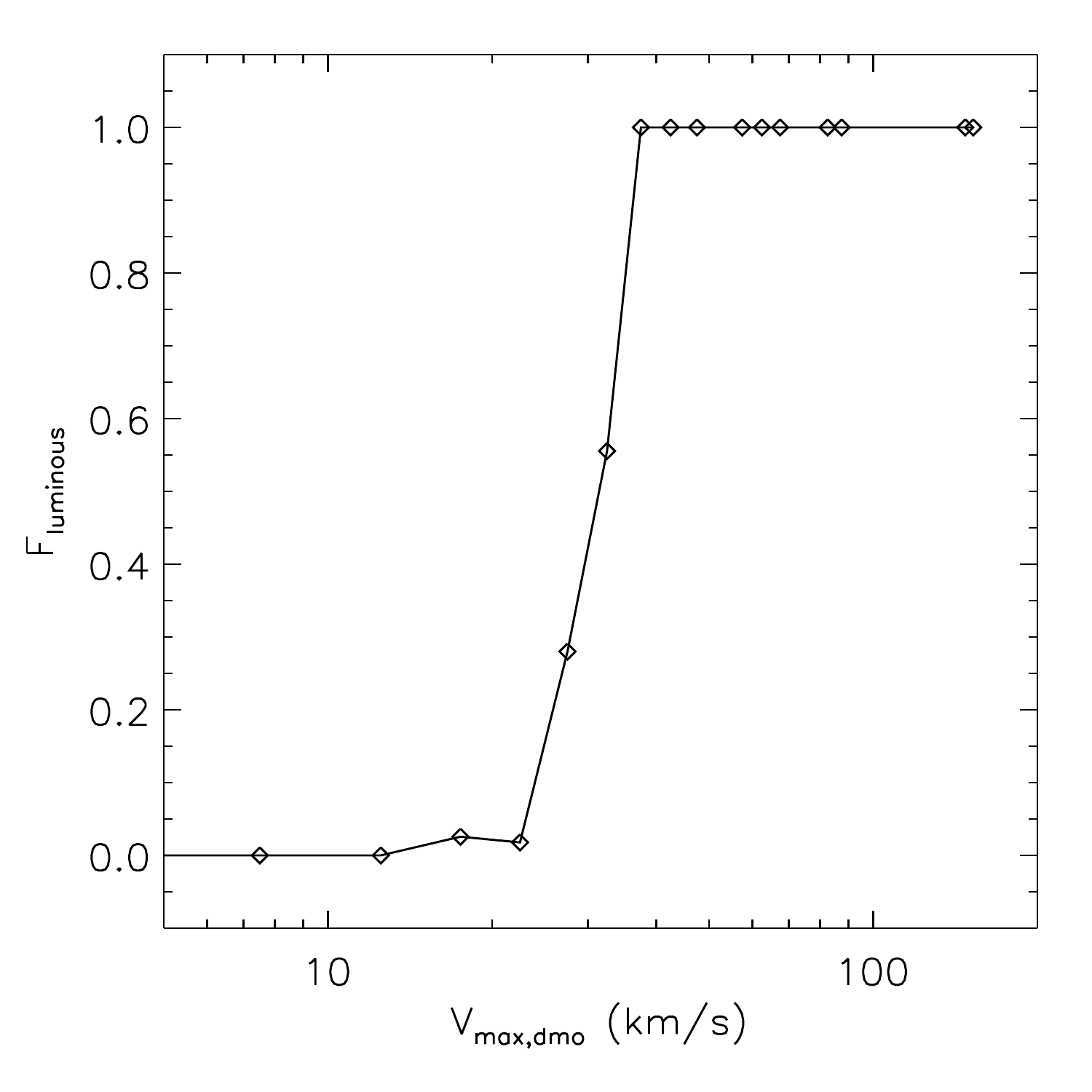}
\caption{The thick solid line represent the fraction of detectable halos as a function of \vmaxdmo \ according to our simulations. The term detectable refers here to halos with $M_\ast > 10^6 \; M_\odot$. This specific stellar mass cutoff was chosen to approximately recover the selection of galaxies used to measure the VF in the Local Volume \citep{Klypin2015}. }
\label{fdet}
\end{figure}

We use this sample to compute the fraction of halos hosting simulated galaxies with $M_\ast > 10^6 \; M_\odot$ in the baryonic runs, as a  function of the maximum circular velocity in the DM-only runs, $f_{det}(v_{max,dmo})$. The $M_\ast > 10^6 \; M_\odot$ cutoff is chosen because it corresponds to the typical stellar mass of galaxies with $M_B = -10$, which define the faint limit of the \citet{Klypin2015} measurement. The result is shown in Figure~\ref{fdet}. As the figure shows, virtually all halos with \vmaxdmo \ $\gtrsim 35$ \kms \ host detectable galaxies, and thus are expected to be included in the VF measurement of \citet{Klypin2015}. On the other hand, the detectable fraction drops precipitously at lower values of \vmaxdmo, falling below the 5\% level at \vmaxdmo \ $\lesssim 25$ \kms. As shown in \S\ref{VF_computation}, this sharp drop in the fraction of detectable galaxies at low values of \vmaxdmo \ has important consequences for the measurement of the low-velocity end of the VF.

Keep in mind that the value of \vmaxdmo \ where the dramatic drop in detectability takes place is dependent on the depth of the galaxy catalog used to measure the VF. If a deeper census of Local Volume galaxies were available, the minimum detectable stellar mass would be lower than $\sim 10^6 \; M_\odot$, and the drop in detectability would consequently appear at lower values of \vmaxdmo \ than shown in Figure~\ref{fdet}. Eventually, a physical effect will limit galaxy formation in halos with very low values of \vmaxdmo, namely reionization feedback \citep[e.g.,][]{Okamoto2008,Sawala2015}.

\subsection{Mock ``observed'' rotational velocities}
\label{velocity_relation}

\begin{figure*}
\plotone{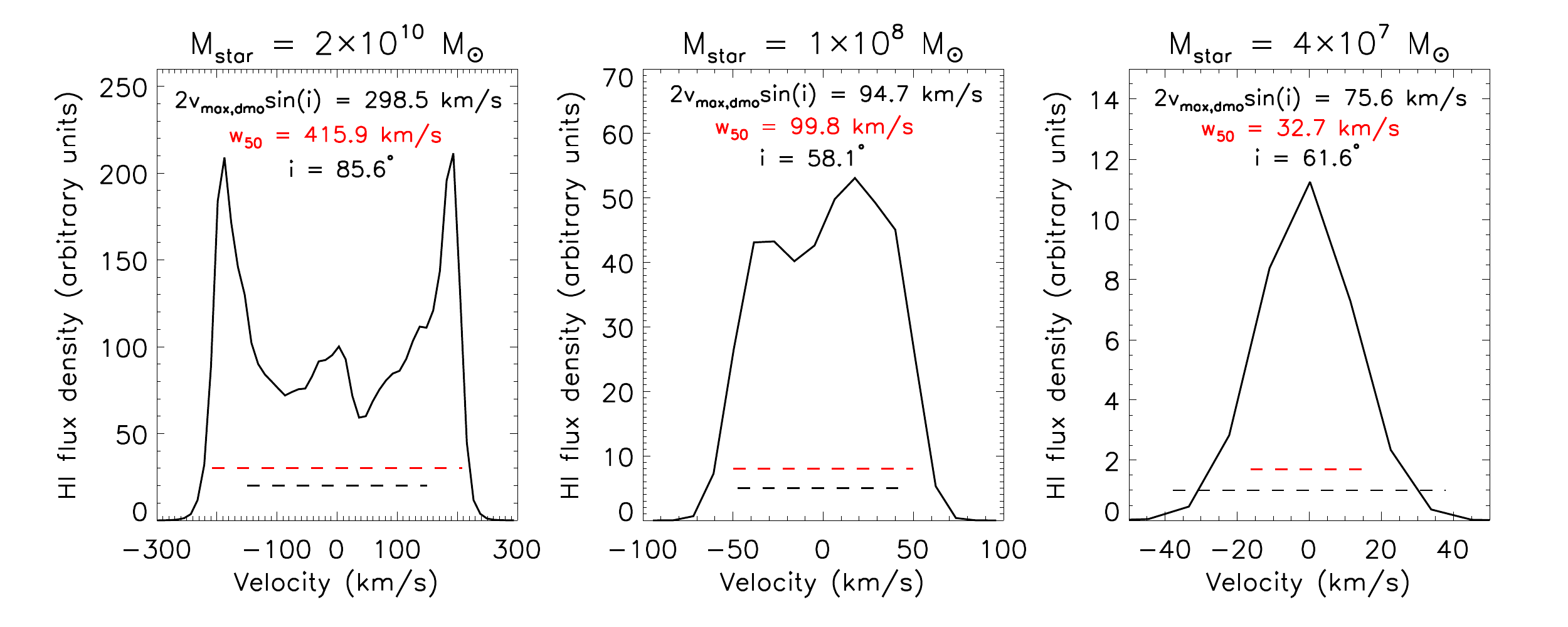}
\caption{
Example HI profiles for three simulated galaxies, viewed at random inclination angles. Their stellar masses are listed at the top of each panel. The 2$v_{max,dmo}sin(i)$ of their DM-only counterparts are printed in each panel, as is the inclination angle, $i$.  2$v_{max,dmo}sin(i)$ is also shown by the black dashed line, while their measured $w_{50}$ at this inclination angle are listed in red and shown as the red dashed line for comparison. Note that the range of both the $x$-axis and the $y$-axis is different in each panel.
}
\label{hiprofiles}
\end{figure*}

Most of the galaxies ($\sim$80\%) in \citet{Klypin2015} have rotational velocities derived from HI. For this reason, we analyze the HI content of our baryonic halos and derive observationally motivated rotational velocities for our simulated galaxies that contain enough HI mass to fall into the \citet{Karachentsev2013} catalog. \citet{Klypin2015} also include dispersion-supported galaxies with no measurable HI down to $M_B = -10$.  In this section, we describe our selection criteria to mimic this sample and derive mock observational velocities.

To restrict our sample to halos with enough baryonic material to fall into the \citet{Klypin2015} sample, we identify all halos in the DM-only zoomed runs that have $v_{max,dmo} \geq 15$ \kms \ at $z = 0$ and their counterparts in the baryonic zoomed runs. 
This yields an initial sample of 57 halos. From this initial sample, we identify those with an HI mass, M$_{\rm HI}$, greater than 10$^6$ M$_{\odot}$, corresponding to the HI mass of the faintest galaxies in the \citet{Karachentsev2013} catalog that have HI linewidth data.  This yields a sample of 42 galaxies with enough HI mass to generate mock HI data cubes (described below).  Of the remaining gas-poor galaxies, we keep only those with $r$-band magnitudes brighter than -10 in the baryonic runs, to approximately mimic the $M_B = -10$ limit of the \citet{Karachentsev2013} catalog. Five out of the initial sample of 57 halos are fainter than this $r$-band cutoff, and are therefore not included in the subsequent analysis. 

Ten gas-poor halos with HI masses below our adopted cutoff, $M_{HI} < 10^6 \; M_\odot$, remain in the sample. Four of these dispersion-supported halos with no HI are satellites of a Milky Way-mass galaxy, and we adopt for them the stellar velocity dispersion as the mock ``observed'' velocity.  For the other six faint galaxies without HI data cubes, we adopted the procedure of \citet{Klypin2015}, who assigned a velocity dispersion of 10 \kms \ to all halos with $M_K$ fainter than -15.5.  Hence, we assign these halos a velocity dispersion of 10 \kms. However, whether we use a fixed 10 \kms \ or the stellar velocity dispersion measured directly from the simulation makes no change to our results, as the simulated velocity dispersions are on the order of 10 \kms, similar to the observational data.  

The HI mass fraction of every gas particle in the baryonic runs is calculated based on the particle's temperature, density, and the cosmic UV background radiation flux, while including a prescription for self-shielding of H$_2$ and dust shielding in both HI and H$_2$ \citep{Christensen2012}. This allows for the straightforward calculation of the total HI mass of each simulated galaxy. We create mock HI data cubes only for the 42 halos that contain M$_{\rm HI} >$ 10$^6$ M$_{\odot}$. Specifically, we create mock data cubes that mimic ALFALFA observations \citep{Haynes2011}. After specifying a viewing angle (see below), our code considers the line-of-sight velocity of each gas particle. The velocity of each particle is tracked in the simulation by solving Newton's equations of motion, but any turbulent velocity of the gas is not taken into account. Velocity dispersions in dwarf galaxies can be on the order of the rotational velocity, $\sim$10-15 \kms \ \citep[e.g.,][]{Stanimirovic2004,Tamburro2009,Oh2015}. Dispersions are thought to be driven at least partially by thermal velocities or supernovae \citep{Tamburro2009, Stilp2013a, Stilp2013b}. In our simulations, supernovae inject thermal energy, and the thermal state of the HI gas needs to be considered in the mock HI linewidth for a realistic comparison to observations.  To account for the thermal velocity, the HI mass of each gas particle is assumed to be distributed along the line-of-sight in a Gaussian distribution with a standard deviation given by the thermal velocity dispersion, $\sigma = \sqrt{kT/m_{HI}}$, where $T$ is the temperature of the gas particle. After this thermal broadening is calculated, a mock HI data cube can be generated by specifying the spatial and velocity resolution.  
For all of our galaxies, we adopted a spatial resolution of 54 pixels across 2$R_{vir}$. In practice, this corresponds to $\sim$1kpc resolution in our lowest mass galaxies to up to $\sim$9kpc resolution in our most massive galaxies.
However, the spatial resolution plays no role in our study, since measurements of the VF are based on spatially unresolved HI data.  For the velocity resolution, we match the ALFALFA specification of 11.2 \kms \ (two-channel boxcar smoothed).

\begin{figure*}
\centerline{\includegraphics[angle=0,width=0.95\textwidth]{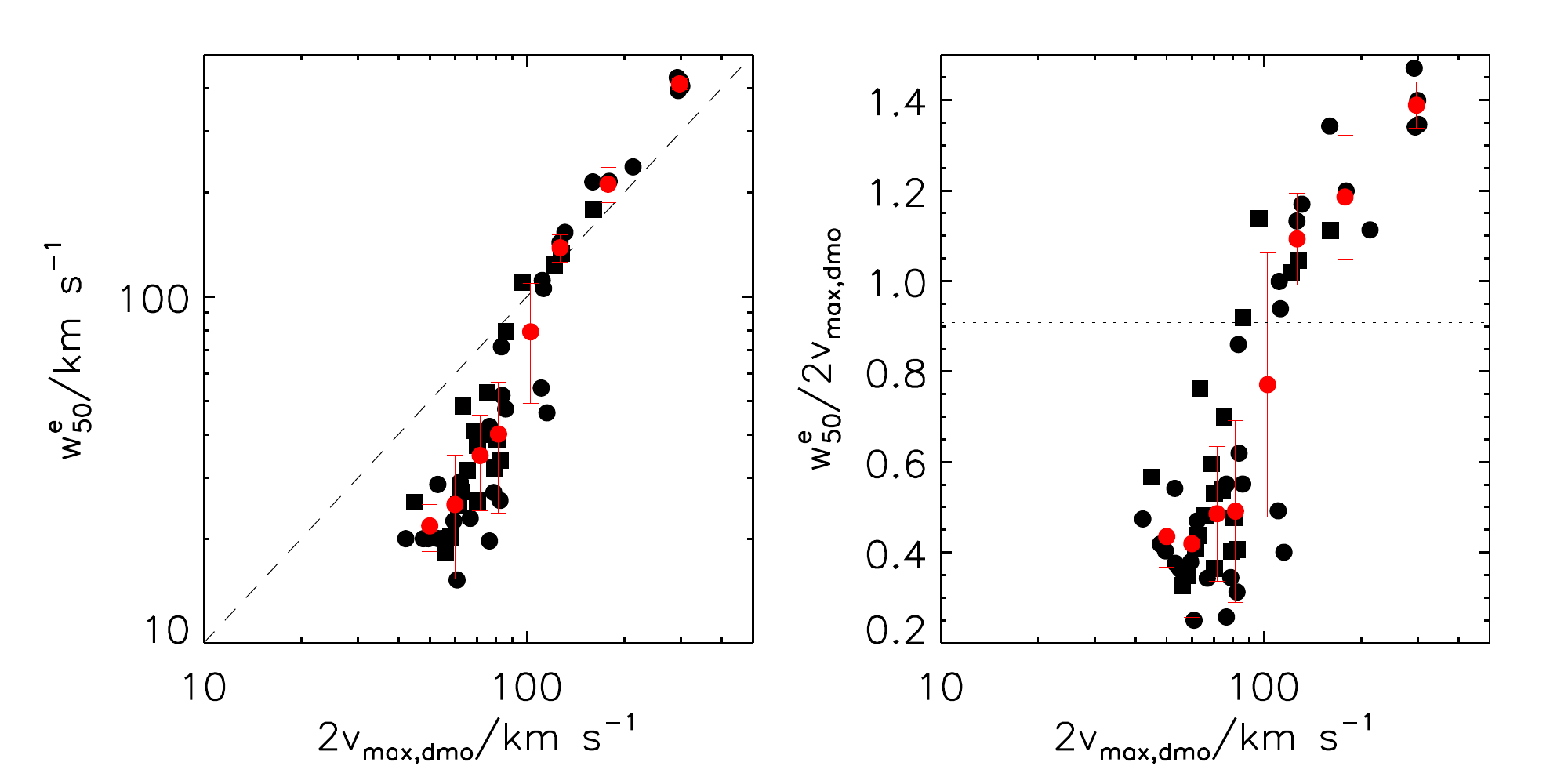}}
\caption{
The measured rotational velocity compared to the theoretically derived expectation.  {\it Left}: The value of $w_{50}$ for the baryonic runs, measured from the edge-on mock HI data cubes, or the stellar velocity dispersion for low mass galaxies without HI, is plotted against 2v$_{max,dmo}$ from the same galaxies run with DM only (black points).  In both panels, the dashed line shows a one-to-one relation, and subhalos are shown as squares while central galaxies are shown as circles. The subhalos show no systematic difference from the central galaxies.  The red points in both panels show the average relation in bins containing four to ten data points, depending on the density of the data.  Error bars reflect the 1$\sigma$ standard deviation about the average. {\it Right}: $w_{50}$/2v$_{max,dmo}$ versus 2v$_{max,dmo}$.  This emphasizes that galaxies with 2v$_{max,dmo} \gtrsim 150$ \kms \ show slightly higher rotational velocities in the baryonic run than in their DM-only counterparts.  At lower velocities, the dwarfs are measured to have a substantially lower $w_{50}$ than 2v$_{max,dmo}$.  The dotted line shows the reduction in circular velocity expected from loss of baryons alone. 
} 
\label{fig4}
\end{figure*}

For each of the 42 galaxies with $M_{HI} > 10^6 \; M_\odot$, we create two HI data cubes. In the first case, we orient each galaxy to be viewed edge-on, i.e., such that the HI angular momentum vector is lying in the image plane. This generates HI data cubes without inclination effects. In the second case, we pick a random orientation of each simulated galaxy (the x-axis of the simulation volume in all cases) and generate HI data cubes that capture inclination effects. In both cases, we measure the width of the HI profile at 50\% of the peak height. Hereafter, we denote the edge-on velocity width by \wfiftye, while we denote the velocity width projected at a random inclination angle by \wfifty. The latter projected velocity width, \wfifty, is the one that can be directly measured observationally. 
For the 10 gas-poor, dispersion-dominated galaxies, we define both \wfiftye \ and \wfifty \ to be twice the stellar velocity dispersion.

Example HI line profiles for three simulated galaxies spanning a large range of mass are shown in Figure~\ref{hiprofiles}. The HI profiles  are derived at random inclination angles, which are indicated in each panel. The figure demonstrates how the HI rotational velocity can differ from the simple theoretical expectation based on the DM-only runs. In particular, we compare the measured \wfifty \ of the simulated galaxies to its simplest theoretical equivalent\footnote{The form of the ``theoretical velocity width'', \twovmaxsini, follows from the fact that the HI profiles plotted in Fig.~\ref{hiprofiles} are projected on a viewing angle of inclination $i$, and include emission from both the approaching and receding sides of the HI disk. In the text we will generally use \twovmax \ to compare DM-only velocities with edge-on HI velocity widths, \wfiftye, and \twovmaxsini \ to compare with projected HI velocity widths, \wfifty.}, $2\,v_{max,dmo}\,\sin i$.  The example simulated galaxies shown in Figure \ref{hiprofiles} demonstrate a trend that has a profound impact on the computation of the theoretical VF. In particular, the HI velocity for massive galaxies is larger than the DM-only velocity, \wfifty$>$\twovmaxsini. This shift to higher velocities in the baryonic run is attributed to the cooling of baryons onto the central halo in massive galaxies \citep[e.g.,][]{Trujillo-Gomez2011, Dutton2011}. On the contrary, low-mass simulated galaxies display the opposite effect. The single-peaked shape of their HI profile leads to a measured value of \wfifty \ that is significantly smaller than \twovmaxsini.

Figure~\ref{fig4} shows the relation between the mock observational and DM-only rotational velocities for all our baryonic halos. More specifically, we compare the edge-on velocity widths, \wfiftye, with the equivalent edge-on DM-only widths, \twovmax. This is done in order to facilitate a direct comparison that neglects inclination effects.  The red points show the average relation in bins containing four (in the highest velocity bins) to nine (in the lowest velocity bins) data points, depending on the density of the data. Error bars reflect the 1$\sigma$ standard deviation about the average.  The dashed line in both panels shows a one-to-one relation between the baryon and DM-only results.  It is obvious from this plot that galaxies with 2v$_{max,dmo} \gtrsim 150$ \kms \ show higher velocities in the baryonic runs than the DM-only runs, while the trend is reversed at lower masses. The dotted line in the right panel shows the decrease expected in velocity from the DM-only runs if all of the baryons had been lost from the halo. The lowest mass galaxies show a much larger change than can be explained due to baryon loss alone.\footnote{Note that even if a simulated dwarf galaxy loses a large percentage of the cosmic baryon fraction it remains gas-rich at $z=0$ due to the fact that the gas that remains behind is inefficient at forming stars, unless it is a satellite and has had its gas stripped.} We dissect the reasons for this lower-than-expected velocity in Section \ref{why}.

Twenty of the 52 halos plotted in Figure~\ref{fig4} are subhalos (denoted by squares) of larger halos. As seen in this figure and those that follow, the simulated galaxies hosted by subhalos follow similar kinematic trends to those hosted by central halos.

\subsection{Re-deriving the Expected VF} 
\label{VF_computation}

\begin{figure}
\includegraphics[angle=0,width=\columnwidth]{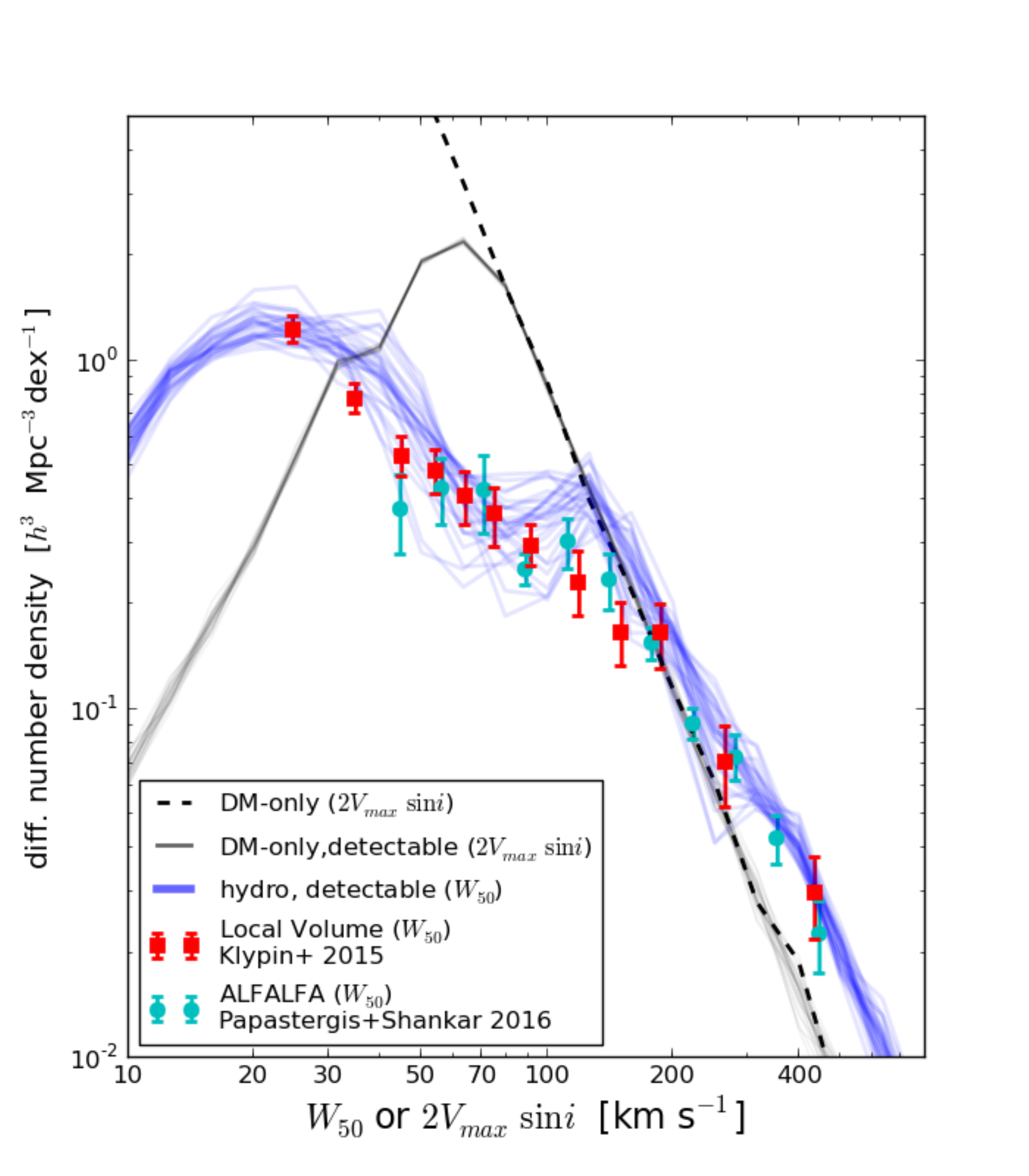}
\caption{
The expected VF including baryonic effects compared to observations. The red and cyan datapoints with errorbars are observational measurements of the VF, based respectively on Local Volume galaxies and on galaxies detected by the ALFALFA survey \citep{Klypin2015,Papastergis2016}. The observational VFs are plotted in terms of $w_{50}$, which is the line-of-sight width of the HI line profile. The black dashed line is the theoretical VF of halos in a DM-only simulation with \textit{Planck} cosmological parameters \citep{Rodriguez2016}. The thin grey lines are the DM-only VF for halos that are expected to host detectable galaxies with $M_\ast > 10^6 \; M_\odot$, according to our simulations (Fig.~\ref{fdet}). These DM-only VFs are plotted in terms of $2v_{max,dmo}$, which is twice the maximum circular velocity of the halo in the DM-only case. The thick blue solid lines are the expected VF of detectable \textit{galaxies} according to our simulations, derived based on the $2v_{max,dmo} - w_{50}$ relation observed for our simulated galaxies (Fig.~\ref{fig4}), and it is corrected for line-of-sight projection assuming random galactic orientations.  The different line bundles represent the uncertainty in the simulation results stemming from uncertainties in the fraction of luminous halos (Fig.~\ref{fdet}) and in the $w^e_{50} -- 2v_{max}$ relation (Fig.~\ref{fig4}) due to the finite number of simulated objects used to calculate them.
}
\label{fig5}
\end{figure}

Based on the results of \S\ref{detectability} and \S\ref{velocity_relation}, we can now compute a realistic expectation for the VF of galaxies in a \LCDM \ universe. The process is illustrated in Figure~\ref{fig5}. In particular, we start from the VF of halos in a \LCDM \ universe with \textit{Planck} cosmological parameters \citep{Planck2014}. This DM-only VF is plotted as a black dashed line in Fig.~\ref{fig5}, and is obtained from the BolshoiP dissipationless cosmological simulation \citep{Rodriguez2016}. The halo VF represents the number density of halos as a function of their maximum circular velocity \vmaxdmo. We denote the theoretical DM-only VF by 

\begin{eqnarray}
\phi_h(v_{max,dmo})= \frac{dN_h}{dV\,d\log_{10}(v_{max,dmo})} \;\;\; .
\end{eqnarray}

\noindent
In the equation above, $dN_H$ is the number of halos contained in a representative volume element $dV$ of the universe, that have rotational velocities within the logarithmic velocity bin $d\log_{10}(v_{max,dmo})$.  

Second, we correct the plotted DM-only halo VF to take into account the detectability of halos as a function of \vmaxdmo. We perform this correction based on the result of Figure~\ref{fdet}. In particular,

\begin{eqnarray}
\phi_{h,det} = f_{det}(v_{max,dmo}) \times \phi_h(v_{max,dmo}) \;\;\; .
\end{eqnarray}

\noindent
The corrected DM-only VF is plotted in Fig.~\ref{fig5} as thin grey lines.  The bundles of lines represent the uncertainty due to the number of simulated halos used to make Figure~\ref{fdet}.

Lastly, we compute the change in the theoretical VF that is due to the difference between the theoretical and observational measures of rotational velocity,

\begin{eqnarray}
\phi_{h,det}(v_{max,dmo}) \rightarrow \phi_{h,det}(w_{50}) \;\;\; .
\end{eqnarray}

This is done by first generating a large number of \vmaxdmo \ values according to the DM-only halo VF corrected for halo detectability (grey lines in Fig.~\ref{fig5}). We then assign to each generated halo an edge-on velocity width value, \wfiftye, based on the mean and scatter of the \wfiftye-\twovmax \ relation shown in Figure~\ref{fig4}. Lastly, we calculate the projected HI velocity width as $w_{50} = w_{50}^e \times \sin i$. Inclination values, $i$, are drawn assuming random orientations, i.e., such that $\cos i$ is uniformly distributed in the $[0,1]$ interval. The final results for the baryonic VF expected in a \LCDM \ cosmology according to our simulations are shown by the blue lines in Figure \ref{fig5}.  The bundles again represent the uncertainty due to the combined uncertainties introduced by the number of simulated halos used to calculate detectability and the number of galaxies in each of the bins in Figure~\ref{fig4}. This distribution is \textit{directly comparable} to the observational VF measured in the Local Volume by \citet{Klypin2015}.

Figure \ref{fig5} clearly demonstrates that taking into account both the ``observed'' velocities and the luminous fraction of halos has a dramatic effect on the theoretical VF. At the high velocity end, the baryonic VF displays a higher normalization than the DM-only distribution, which is caused by the fact that the HI velocity width, \wfifty, is larger than \twovmaxsini \ for massive halos (refer to Figs.~\ref{hiprofiles} \& \ref{fig4}; though note the effect appears less strong in Figure~\ref{fig5} because it shows 2\vmaxdmo \ instead of \twovmaxsini). However, baryonic effects have their largest impact on the low-velocity end of the theoretical VF. In particular, the fact that low-mass halos have \wfiftye \ values significantly smaller than than 2\vmaxdmo \ means that the theoretical VF systematically ``shifts'' towards lower velocities in the dwarf regime. This translates into a substantial reduction of the VF normalization at $w_{50} \lesssim 100$ \kms.

At even lower velocities, $w_{50} \lesssim 40$ \kms, the very low detectability of small halos further suppresses the normalization of the baryonic VF. Together, the effects of the baryonic velocity shift and of halo detectability lead to a dramatic decrease on the number of low-velocity galaxies expected in \LCDM, compared to the simplistic DM-only estimate. As Fig.~\ref{fig5} shows, the difference is more than an order of magnitude already at $w_{50} = 50$ \kms. This huge suppression in the number density at low velocities brings our theoretical VF in agreement with the observational measurements, and shows no signs of the overproduction of dwarf galaxies typically encountered in \LCDM.

\begin{figure}
\includegraphics[angle=0,width=\columnwidth]{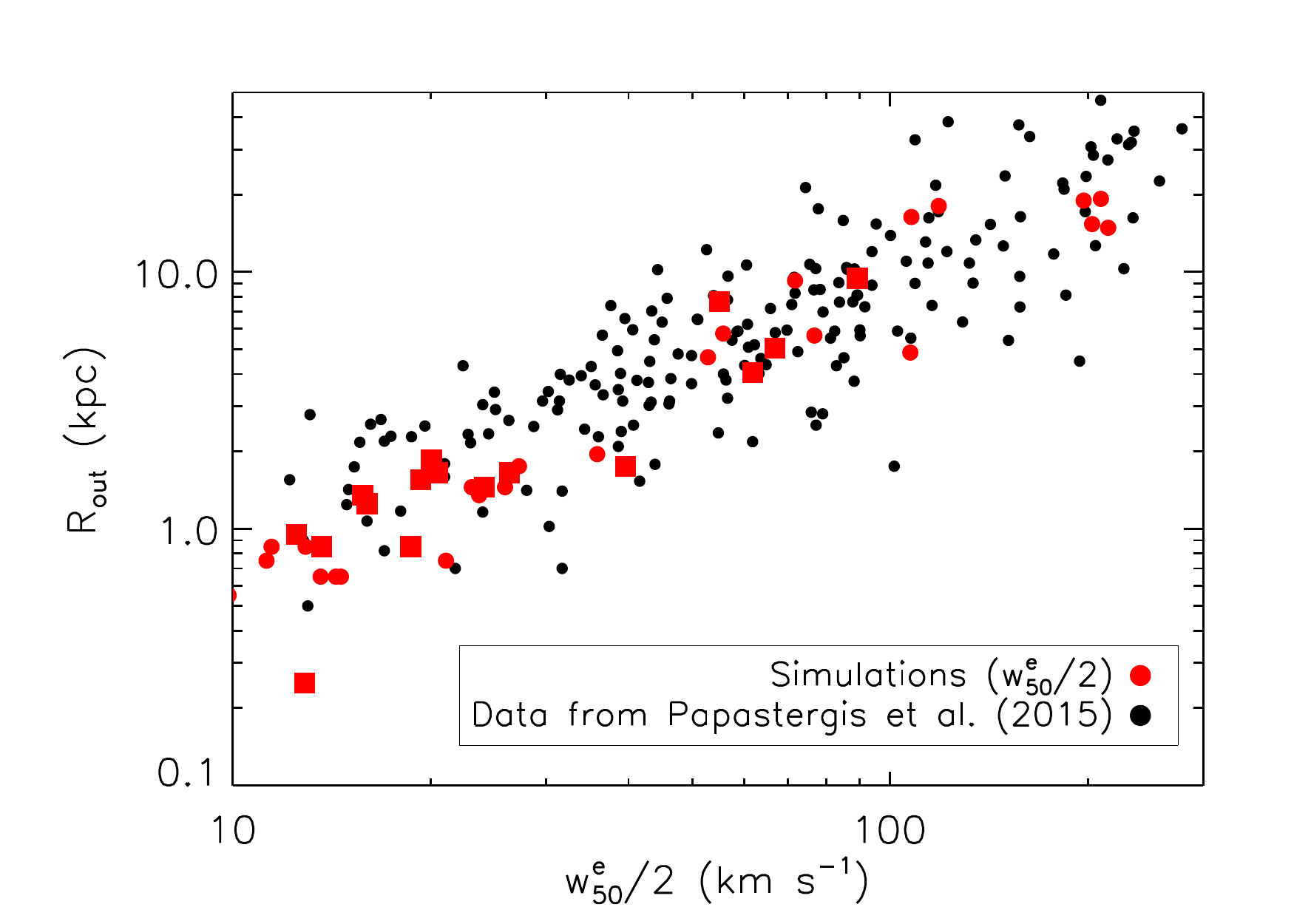}
\caption{
The outermost HI radius as a function of galaxy rotation velocity, $w^e_{50}$. Black data points are observational data from \citet{Papastergis2015},  and the red points are from our baryonic simulations, and measure the radius at which the HI surface density falls below 1 M$_{\odot}$/pc$^2$. Red squares are simulated subhalos, while red circles are central galaxies. Simulations and data follow a similar trend, i.e., simulations and observations are tracing a similar radius as a function of measured velocity. 
}
\label{rout}
\end{figure}

\subsection{Validation Against Other Scaling Relations}

A key point regarding the results of Fig.~\ref{fig5} is that reproducing the observational VF in a simulation is not physically meaningful unless the typical HI disk sizes in dwarf galaxies are also reproduced correctly. 
This is because the ratio between \wfiftye \ and \twovmax \ can be made arbitrarily small in dwarf galaxies by producing simulated galaxies with very small HI disks. 
Because the innermost portion of the rotation curve is rapidly rising, 
it could be possible to reproduce the observed VF but not accurately reproduce observed disk sizes. 
Figure~\ref{rout} compares the sizes of HI disks in our simulated galaxies with the observed sizes in the sample of galaxies with interferometric HI observations compiled by \citet{Papastergis2015}. The observational datapoints show the outermost radius where the HI rotational velocity can be measured by the interferometric observations for each galaxy. One complication here is that the outermost HI radius for the galaxies in the \citet{Papastergis2015} sample is not defined in a consistent way, but depends on the depth of each interferometric observation and the quality of each galaxy's kinematics. For the simulations, we derive ``outermost'' HI radii where the HI surface density profiles of our simulated galaxies fall below 1 M$_{\odot}$/pc$^2$. The adopted HI surface density cutoff corresponds to the value probed by typical interferometric HI observations. We examined the results using different definitions of ``outermost'' HI radius for our simulated galaxies, and found that the results were generally consistent but that this definition produces the least scatter. This is not surprising, because we have also verified that our simulated galaxies follow the observed HI mass -- radius relation from \citet{Wang2016}, where the HI radius is again defined at the 1 M$_{\odot}$/pc$^2$ isophote. The observed relation has remarkably low scatter, so it is reassuring that using a similar definition for the simulations also produces the smallest scatter. Overall, Fig.~\ref{rout} shows that our simulated galaxies have HI disk sizes that are in agreement with observations, indicating that the mock observational velocities computed in \S\ref{velocity_relation} are realistic.

\begin{figure}
\includegraphics[angle=0,width=\columnwidth]{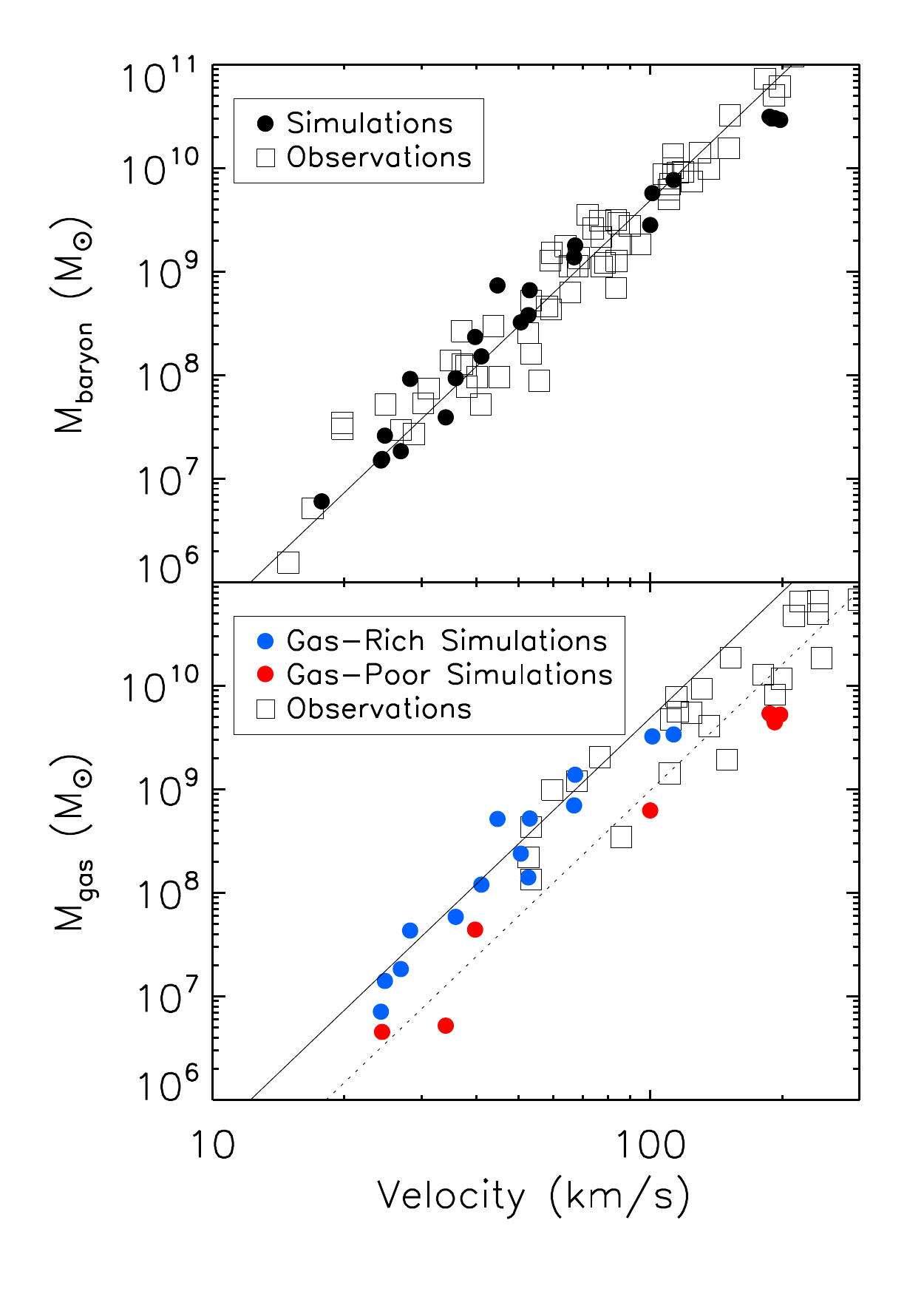}
\caption{The Tully Fisher relation for galaxies.  {\it Top}: The baryonic Tully-Fisher relation (cold gas plus stellar mass). Observed galaxies (squares) from \citet{McGaugh2015} and the best fit relation (solid line).  Simulated central galaxies are shown as black circles and reproduce the baryonic Tully Fisher relation over the sampled velocity range. Observations use the velocity on the flat part of the rotation curve, $v_{flat}$.  We adopt the circular velocity at 4 disk scale lengths as a proxy for $v_{flat}$ in the simulated galaxies. {\it Bottom}: Solid line is the best fit as in the top panel, but the observational and simulation data now only considers cold gas mass (1.33*M$_{HI}$).The simulated sample is divided into a gas-rich (M$_{star}$/1.33M$_{HI} > 2.0$, blue points) and gas-poor (M$_{star}$/1.33M$_{HI} < 2.0$, red points) sample. Observationally, only the gas-rich galaxies follow the full baryonic Tully Fisher relation from the top panel, as do our gas-rich simulated galaxies.  Gas-poor galaxies fall below the observed relation. The dotted line shows a reduction of the relation by a factor of 5.  The match between our simulated sample and observed galaxies as a function of velocity and gas fraction allows us to undertake the study in this paper.}
\label{tf}
\end{figure}

Similarly, the fraction of detectable halos computed in \S\ref{detectability} is not physically meaningful unless our simulations reproduce the baryonic content of real galaxies. In Figure~\ref{tf} we show the baryonic (cold gas plus stellar mass) Tully-Fisher relation for the simulated galaxies used in this work (black points, top panel). We restrict ourselves to central galaxies only (excluding subhalos) for comparison to the observational data, which is taken from \citet{McGaugh2015}. The line in both panels is the baryonic Tully Fisher relation fit to observed galaxies in \citet[][their figure 6 and table 5]{McGaugh2015}, $log(M_b) = 1.61 + 4.04V_f$. Since the \citet{McGaugh2015} measurement refers to the flat outer velocity of galactic rotation curves, we adopt for the simulations the circular velocity of the baryonic runs measured at 4 disk scale lengths as $V_f$. The bottom panel is for the cold gas mass (1.33*M$_{HI}$ in the simulations) only. The simulations have been divided into a gas-rich (M$_{star}$/1.33M$_{HI} > 2.0$, blue points) and gas-poor (M$_{star}$/1.33M$_{HI} < 2.0$, red points) sample. Like the observational data, gas-rich galaxies follow the observed baryonic Tully Fisher relation, while gas-poor galaxies lie below the relation (the dotted line shows a reduction of the relation by a factor of 5). This plot demonstrates that our simulated galaxies match the stellar and HI masses of galaxies as a function of velocity.

Overall, Figures \ref{rout} \& \ref{tf} give us confidence that the theoretical VF computed in \S\ref{VF_computation} is physically well motivated. Consequently, moving from predictions based on DM-only runs to baryonic simulations may be the key to reconciling the theoretical expectation of the VF with the observational measurements.

\section{Velocity Changes in the Presence of Baryons}\label{why}

In this section, we examine the baryonic effects that lead to dwarf halos being observed at lower velocities than predicted based on DM-only simulations, and that help to reconcile the theory with the observations.

It is well known that the rotation curves of many dwarf galaxies are still rising at their outermost measured point \citep{Catinella2006, deblok2008, Swaters2009, Oh2011}, suggesting that the true $v_{max}$ of the halo is higher than HI measures.  In this section we 
use the velocity at the outermost HI data point in our baryonic simulations in order to determine how much of a role this plays in the lowered velocities we see in the dwarf simulations compared to their DM-only v$_{max,dmo}$ values.  Recall that in Figure \ref{rout} we defined the outermost HI data point, R$_{out}$, in our simulations to be the point at which the HI surface density falls below 1 M$_{\odot}$/pc$^2$.  In what follows, we refer to the circular velocity at R$_{out}$ as $v_{out}$.

\begin{figure}
\includegraphics[angle=0,width=\columnwidth]{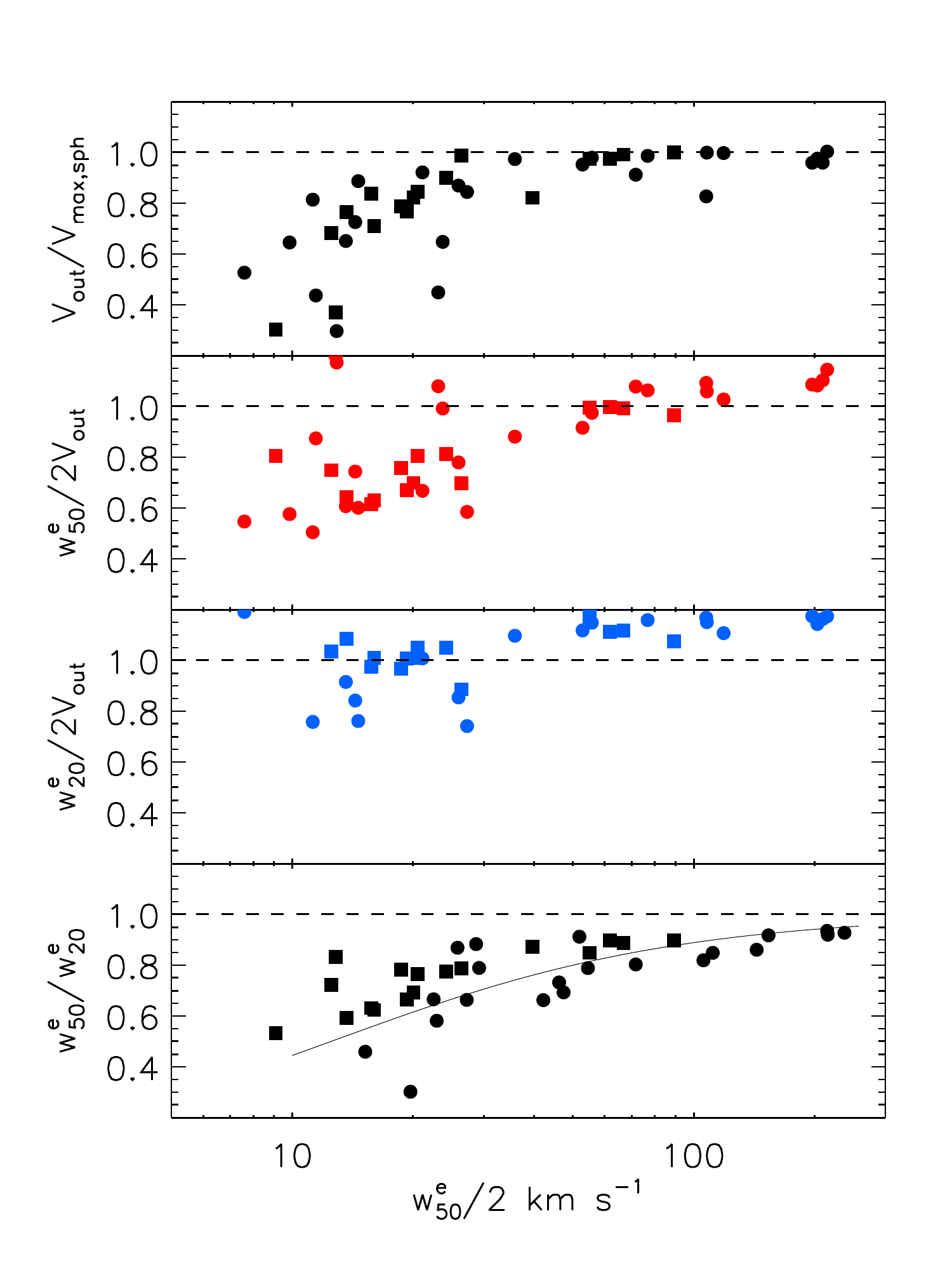}
\caption{Comparison of various velocity measurements.  In all panels, subhalos 
of larger galaxies are shown by squares, while central galaxies are shown by 
circles.  The two groups show similar trends. {\it Top}: The velocity 
measured at the outermost HI data point, $v_{out}$, compared to the maximum 
circular velocity of the baryonic rotation curve, $v_{max,sph}$.  Above $\sim$50 \kms, $v_{out}$ traces 
the rotational velocity.  Below $\sim$50 \kms, however, $v_{out}$ underpredicts 
the maximum rotation speed, due to the fact that the rotation curve is still 
rising at the outermost HI data point. 
{\it 2nd Panel}: $w^e_{50}$ compared to $v_{out}$ as a function of maximum rotational velocity.  If 
$w^e_{50}$ captures the rotational velocity at the outermost HI data point, the 
data points in this panel (red) should be $\sim$1.  Below $\sim$50 \kms, $w^e_{50}$ 
does {\it not} trace the outermost HI rotational velocity.  It tends to be 
systematically lower.  {\it 3rd Panel}: $w^e_{20}$ compared to $v_{out}$ as a 
function of maximum rotational velocity (blue points). Similar to the panel above 
it, but using $w^e_{20}$ rather than $w^e_{50}$.  $w^e_{20}$ does a better job of reproducing the velocity at the outermost HI data point. {\it Bottom}: The ratio of $w^e_{50}$ to $w^e_{20}$ as a function of $w^e_{20}/$2.  It is apparent from this panel that $w^e_{20}$ can measure a substantially larger velocity than $w^e_{50}$ in the dwarfs.  The black line 
shows the relation derived from observational data, $w^e_{20} = w^e_{50} + 25$ km s$^{-1}$ \citep{Koribalski2004, 
Bradford2015}. See text for discussion.}
\label{fig9}
\end{figure}

In our dwarf galaxies, the radius of the outermost HI data is generally still on the rising part of the 
rotation curve.  We quantify this in the top panel of Figure~\ref{fig9}, where 
we compare $v_{out}$ to the maximum value of the circular velocity in the baryonic 
run, $v_{max,sph}$.\footnote{Note that up until now the $v_{max}$ we have been 
dealing with comes from the DM-only runs, $v_{max,dmo}$.  $v_{max,sph}$ will differ 
from from $v_{max,dmo}$ due to processes like baryonic contraction at high masses, or 
loss of most of the baryons from the smallest mass halos.  We wish to quantify how 
well HI traces the rotation velocity after these other factors have had their 
influence, and ultimately determine how well $w_{50}$ is tracing the outermost 
HI rotation velocity.  Hence, we switch to $v_{max,sph}$ in Figure~\ref{fig9}.}
In the more massive galaxies, $v_{out}$ is indeed 
capturing the maximum value of the rotation curve.  However, in galaxies below 
$\sim$50 \kms, the outermost HI rotation velocity systematically underestimates 
$v_{max,sph}$.  

Next we wish to know if \wfiftye ~is tracing the outermost velocity, $v_{out}$. The second panel in Figure~\ref{fig9} shows the ratio of the two.  In the four most massive galaxies, \wfiftye ~traces a slightly higher velocity than the outermost HI rotation velocity, due to the fact that these galaxies have large bulges and higher velocities near their center.  More importantly for interpreting dwarf galaxy data, \wfiftye ~is systematically smaller than $v_{out}$.  In other words, $v_{out}$ is already under-measuring the maximum rotational velocity of the galaxy because it is on the rising rotation curve, but \wfiftye ~is measuring an even lower velocity.  This suggests that \wfiftye ~may be measuring a velocity even closer to the center than $v_{out}$.

Evidence that this is the case is found in the third panel of Figure~\ref{fig9}, where we compare $w^e_{20}$ to $v_{out}$ instead.  $w^e_{20}$ measures the width of the HI profile at 20\% of the peak height rather than 50\%.  While it slightly overestimates $v_{out}$ in galaxies above $\sim$50 \kms, it does a much better job of capturing $v_{out}$ in the lower mass galaxies.  In summary, it seems that $w^e_{20}$ is a more reliable indicator of the outermost measurable rotation velocity in the dwarf galaxies.

%\citep[see also][]{Koribalski2004, Avila-Reese2008, Bradford2015}.

Finally, the bottom panel of Figure~\ref{fig9} shows the ratio between our 
$w^e_{50}$ and $w^e_{20}$ measurements, and demonstrates that $w^e_{20}$ can measure a much larger velocity in the dwarfs than $w^e_{50}$, up to a factor of two larger in the lowest mass galaxies.  This difference has been noted previously.  Using 
ALFALFA data, \citet{Bradford2015} showed that the difference between the two velocities is well described by the relation 
$w^e_{20} = w^e_{50} + 25$ km s$^{-1}$ \citep[see also][]{Koribalski2004}. This relation is shown as the black line in the bottom panel of Figure~\ref{fig9}.  \citet{Brook2016b} showed that the discrepancy between $w^e_{20}$ and $w^e_{50}$ could lead to substantial differences in the slope of the baryonic Tully Fisher relation, while \citet{Brook2016} demonstrated that the use of $w^e_{50}$ instead of $v_{max,dmo}$ could fully explain the difference in the theoretical VF compared to observations. We note that almost all observational measurements of the VF are based on $w^e_{50}$ rather than $w^e_{20}$ \citep{Zwaan2010, Papastergis2011, Klypin2015} due to the fact that it can be hard to measure the line width at 20\% of the peak height due to spectrum noise at typical signal-to-noise ratios.

The change between $w^e_{20}$ and $w^e_{50}$ is likely due to the shape of the HI profile as a function of mass.  As was seen in Figure~\ref{hiprofiles}, more 
massive galaxies exhibit a double-horned profile.  The horns are built up due 
to the piling up of velocity along the flat part of the rotation curve in large 
spirals.  However, lower mass galaxies 
are usually still rising at the outermost HI data point, as discussed above.  
This leads to an HI profile that is more Gaussian.  The drop-off at the edges 
of the double-horned profile is rapid, so that the difference between $w^e_{20}$ and $w^e_{50}$ is small.  However, the Gaussian shape in the dwarfs ensures that this is no longer true.  Measuring lower in the HI profile can lead to a much 
larger velocity width.  These higher velocities must come from further out on 
the rotation curve.  

In summary, the maximum rotational velocity traced by HI does not generally trace the full $v_{max,sph}$ for dwarf galaxies below $\sim$50 \kms.  This is due to the fact that the outermost HI is still on the rising part of the rotation curve.  Additionally, $w^e_{50}$ does not measure the the outermost HI rotation velocity in dwarf galaxies, compounding the problem further.  The combination of these two effects leads to the shift in velocities measured between the baryonic and DM-only simulations seen in Figure \ref{fig4}.

\section{Does dark matter core creation matter?}\label{corecreation}

Recent high resolution cosmological simulations of galaxies, including those 
used in this study, have shown 
that feedback from young stars and supernovae can create dark matter cores 
in galaxies \citep[e.g.,][]{Governato2010, Teyssier2013, Chan2015, Dutton2016}. \citet{Governato2012}   
and \citet{diCintio2014a} showed that this result varies with stellar mass (and 
thus also with halo mass, given that there is a stellar-to-halo mass relation).  
The shallow potential wells of dwarf galaxies at $M_{vir} \sim 10^{10} M_{\odot}$ 
are particularly susceptible to core creation, but the deeper potential wells 
of MW-mass galaxies are less so, and galaxies have a harder time creating large cores in lower mass halos that form less stars and therefore inject 
less energy \citep{Maxwell2015, Read2016a}.  

In this section we explore whether the change in the dark matter profile in 
dwarf galaxies has any impact on the observed VF.  Work by \citet{Brook2015} concluded that measuring a theoretical velocity at the radius which reproduces \wfiftye ~is not enough to match observed velocities in models that retain a cuspy, NFW dark matter density profile.  Instead, they showed that additionally considering dark matter core creation could lower the theoretical velocities enough to bring them in line with observations.  We demonstrate here that this is true only for galaxies which have R$_{out} \lesssim 3$kpc.

\begin{figure}
\includegraphics[angle=0,width=\columnwidth]{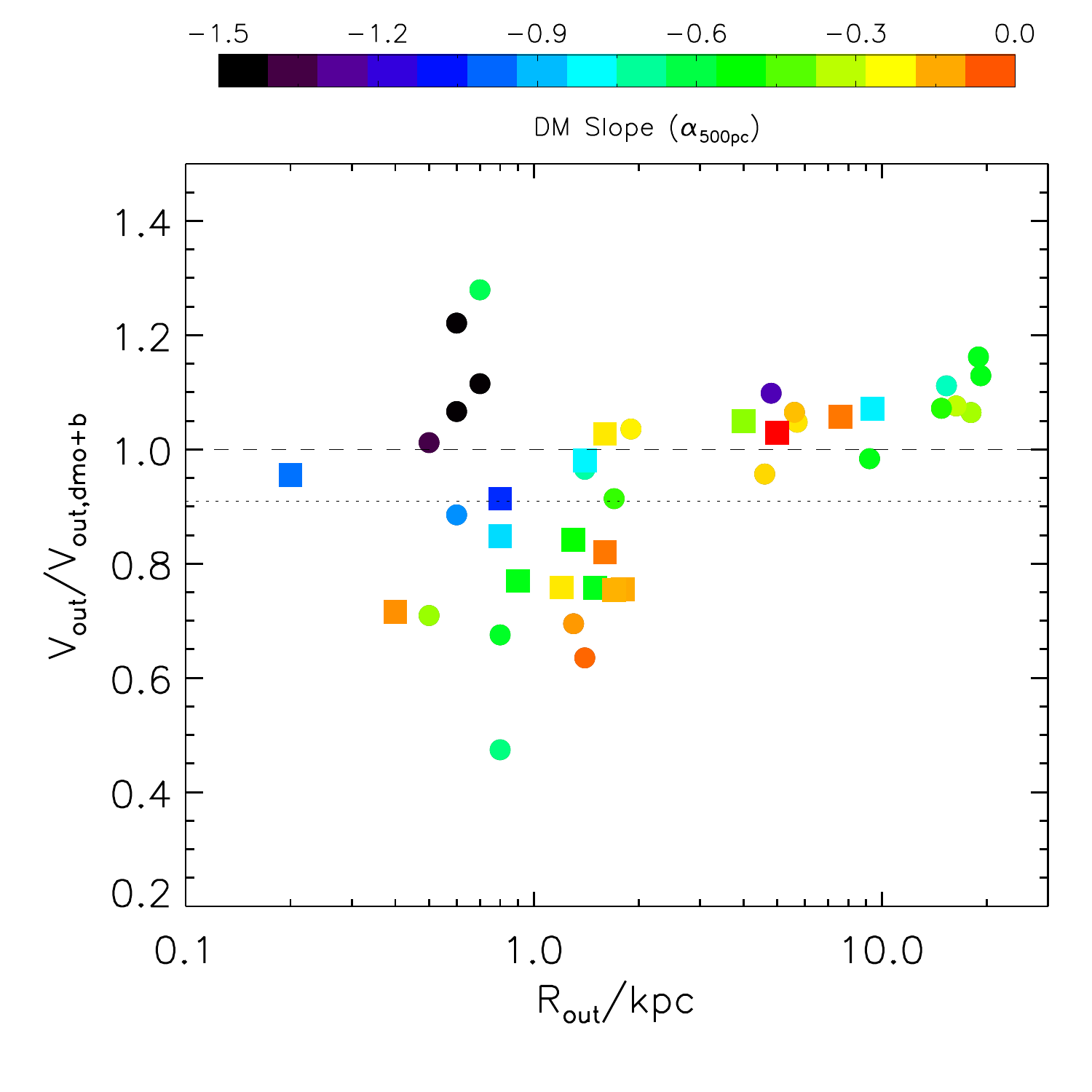}
\caption{Testing the role of dark matter cores. 
The ratio of $v_{out}$ to the velocity at the same radius in the DM-only run but including the baryonic potential, $v_{out,dmo+b}$ (see text for details), as a function of R$_{out}$. The points are color coded by the slope of the dark matter density profile in the baryonic galaxy, according to the color bar at top. 
For galaxies with R$_{out} < 2-3$ kpc, cored galaxies generally have a lower velocity than their cuspy counterparts. 
Circles are field galaxies, squares are subhalos.  All simulated galaxies with HI are included.  Dashed lines are a ratio of 1, while dotted lines represent the ratio expected if velocities are suppressed solely due to a loss of baryons.
  }
\label{cores}
\end{figure}

%\subsection{The Impact of Cores on Measured Velocity}

Assessing the impact of core creation is not simple because the densities in the baryonic simulations may also be subjected to some level of contraction due to the presence of the baryons, and disentangling the two effects is not straightforward. Note that this contraction does not have to be adiabatic contraction of the dark matter, and in fact adiabatic contraction of the dark matter is unlikely to occur in the dwarf regime that we are exploring here.  However, as we demonstrate below, the fact that gas can cool to the center of the galaxy can increase the rotation velocity in the inner regions, even in dwarf galaxies, in the baryonic simulations.  This effect must be accounted for before a direct comparison can be made between the velocities in the baryonic runs and the DM-only runs.  If it is not accounted for, a comparison between the baryonic and DM-only velocities would minimize the impact of dark matter core creation.

To overcome this, we develop a proxy for a contracted model without core creation by adding together the velocity profile in a DM-only run\footnote{A DM-only simulation contains the cosmic density of all matter, $\Omega_{baryon}$ and $\Omega_{DM}$. Here, we scale down the velocity profile of the DM-only run by an amount consistent with removing the cosmic baryonic fraction, so that we can add the baryonic contribution from the SPH runs instead.} with the velocity profile of only the baryonic component in its counterpart SPH run.  This effectively ``contracts'' the profile due to the presence of baryons, but does not include dark matter cores since the DM-only runs do not experience core creation.  We measure the velocity from this combined model at the outermost HI radius, R$_{out}$, determined from the SPH runs, and label it $v_{out,dmo+b}$.

In Figure~\ref{cores} we compare $v_{out}$ measured in the SPH runs to $v_{out,dmo+b}$ as a function of R$_{out}$.  The ratio $v_{out}/v_{out,dmo+b}$ gives us an estimate of how much core creation alone has supressed the rotation curve in the baryonic runs. The data points are color coded based on the slope of their dark matter density profile, measured between 300-700pc, labeled $\alpha_{500pc}$.  We include in this plot all simulated galaxies with HI.  The lowest mass galaxies have star formation efficiencies too low to create substantial dark matter cores.  Core creation is not the only mechanism that can suppress the rotation curve, as loss of baryons alone can lower the baryonic rotation curve relative to the DM-only case.  The ratio expected for pure baryonic mass loss is shown by the dotted line in Figure~\ref{cores}. If core creation is important we would expect to see that the strongly cored galaxies lie systematically lower than other galaxies.  We find this is only true for galaxies with R$_{out} <$ 2-3 kpc. 

For galaxies with R$_{out} < 3$ kpc (corresponding to $v_{out} < 50$ \kms), the galaxies with dark matter cores generally occupy the lowest velocity ratios.  This suggests that core creation contributes to velocity suppression in this regime. The velocities can be lower by up to 40\%, comparable to the reduction from measuring on the rising part of the rotation curve alone (see top panel of Figure~\ref{fig9}).  Thus, cores do seem to substantially contribute to lowered velocities for galaxies with R$_{out} < 3$ kpc.

\begin{figure}[htbp]
\includegraphics[angle=0,width=\columnwidth]{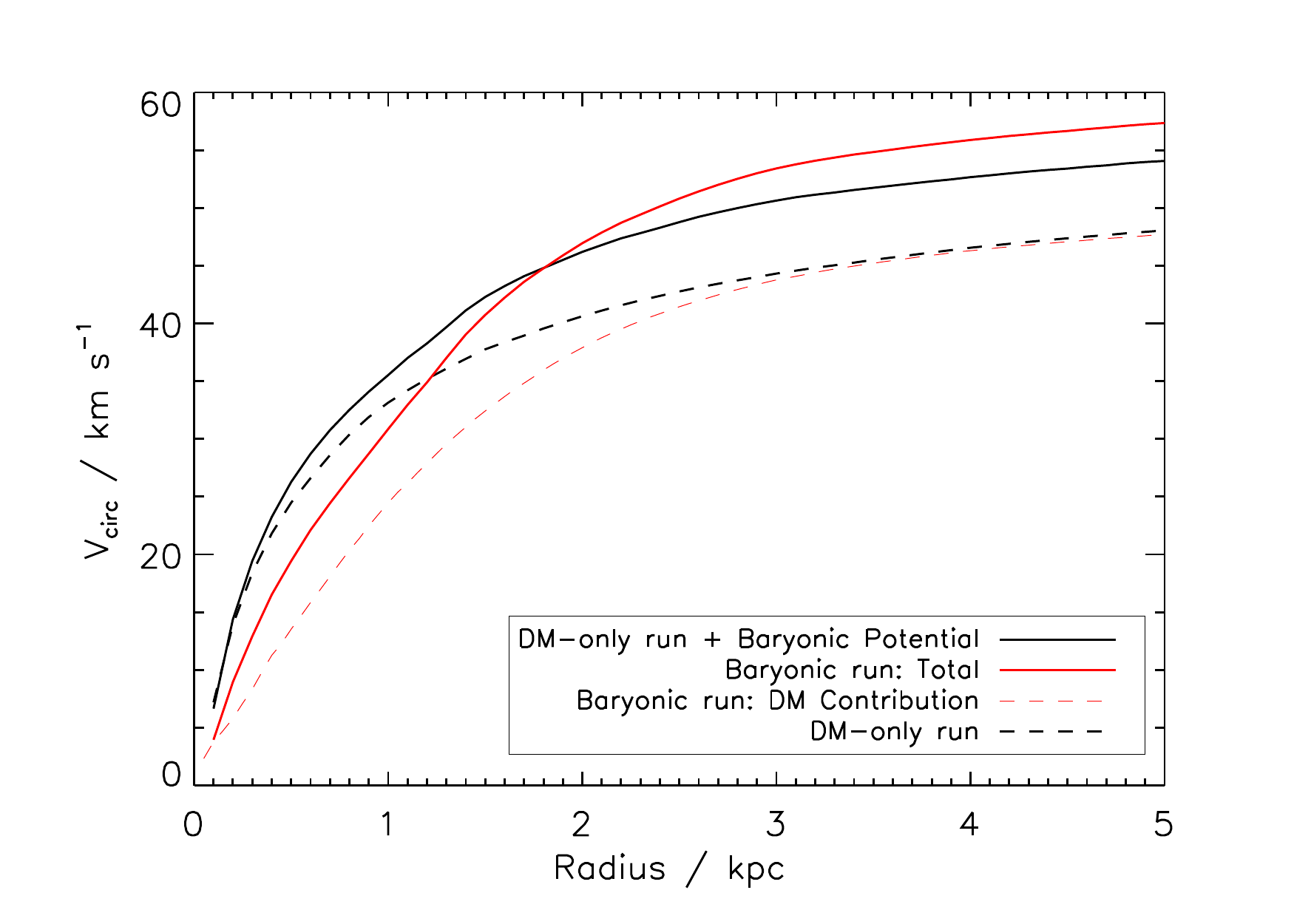}
\caption{Example rotation curve for a dwarf galaxy.  The baryonic run of 
this galaxy has a cored DM profile, as is evidenced by the more slowly 
rising rotation curve (solid red line) in the central region compared to 
the DM-only run (black dashed line) or the combined DM-only/baryonic model (black solid line).  The red dashed line shows the DM 
contribution to the baryonic rotation curve.  Although truncated in this 
plot, in the DM-only run $v_{max,dmo} = 55.8$ \kms \ (which it reaches at 27 kpc), while the baryonic run reaches $v_{max,sph} = 58.3$ \kms \ at 7.5 kpc, i.e., the two runs have comparable $v_{max}$.  The ability of gas to cool puts more mass in the central region of the halo in the baryonic run, so that it reaches $v_{max}$ at a smaller radius than the DM-only run.} 
\label{h516}
\end{figure}

For galaxies with R$_{out} > 3$ kpc, there are strongly cored galaxies that do not show any signs of having their velocities reduced.  In Figure~\ref{h516}, we provide an example of why a galaxy with a strong dark matter core may not have a lower velocity. 
Figure~\ref{h516} shows the rotation curve for one of our dwarf galaxies that 
undergoes significant dark matter core creation. At $z=0$ this halo has a dark matter density slope of -0.3.   
This profile causes the rotation curve to rise much more slowly in the baryonic run (red solid line) compared to the combined DM-only/baryonic model (solid black line) or the DM-only run (black dashed line).  The red dashed line shows the DM contribution to the baryonic run's total $v_{circ}$, to emphasize the presence of the dark matter core.

It can be seen that the baryonic run has a lower rotational velocity than the 
combined DM-only/baryonic model interior to $\sim$2 kpc.  It is clear from Figure~\ref{h516} that if the HI is tracing velocity interior to $\sim$2kpc, then core creation would reduce the measured velocity in this galaxy.  However, this galaxy has HI gas that extends out to roughly 5 kpc, where it is tracing the flat part of the rotation curve, and is an excellent measure of $v_{max,sph}$.  

This galaxy also highlights another subtle point.  
The DM-only run reaches a $v_{max,dmo} = 55.8$ \kms \ at 27 kpc.  The baryonic run reaches $v_{max,sph} = 58.3$ \kms \ at 7.5 kpc.  The velocity from the HI profile, $w^e_{50}$, is 55 \kms, comparable to the $v_{max,dmo}$ measured in the DM-only run.  Thus, there is almost no change in $v_{max}$ between the two runs, i.e., this halo does not undergo adiabatic contraction in the usual sense.  It is simply that the radius at which $v_{max}$ occurs is quite different.  In the baryonic run, the fact that gas can cool leads to the mass being more centralized than in the DM-only run, without increasing $v_{max}$ overall.  Likewise, the ``contracted'' model combining the DM-only profile with the baryonic profile is not adiabatically contracted, but simply reaches $v_{max}$ at a smaller radius.  The cold gas increases the central velocity relative to the DM-only run despite 
the fact that this dwarf is dark matter dominated overall, with a baryon ratio (cold gas and stellar mass to total DM mass) of only 2\% at $z = 0$ (but remains gas-rich due to the fact that star formation is inefficient).

There are a total of four galaxies in our sample where the DM-only counterpart has $v_{max,dmo} \sim$ 55 km s$^{-1}$, like the galaxy shown in Figure~\ref{h516}.  All of these galaxies have stellar masses between 1.5-3$\times$10$^8$ M$_{\odot}$, and all have a cored dark matter density profile, but their HI masses vary by an order of magnitude.  Two of them have R$_{out} \sim$ 1.5 kpc, while two have R$_{out} \sim$ 5 kpc.  As expected, the two with small R$_{out}$ have substantially lower $w^e_{50}$ values compared to $v_{max,dmo}$. Hence, scatter in the HI content at a given halo mass leads to scatter in the role of dark matter cores.

From these examples, we learn that if core creation is to impact the measured 
velocity in a galaxy, the HI must not extend significantly further than the size of the dark matter core. A similar conclusion was found by \citet{Papastergis2016} by analyzing observational dwarf data. In simulated galaxies with efficient core creation, the dark matter cores are often 1-2 kpc.  From Figure~\ref{cores}, we see that the strongly cored galaxies with R$_{out} < 2$ kpc do indeed tend to show a lower rotation velocity in the baryonic run than their DM-only counterpart.

\section{Comparison with Previous Works}

In this section we discuss how our results compare to previous works on this topic.  First we focus on the ability to reproduce the VF, then specifically on the impact of dark matter cores.

\subsection{Velocities}

\citet{Brook2016} were the first to show explicitly that using $w^e_{50}$ instead of $v_{max,dmo}$ could reconcile the theoretical VF with the observed VF.  Their approach was semi-empirical, using abundance matching (a relationship between baryonic mass and halo mass) convolved with a relation between between baryonic mass and velocity.  They showed the impact of using various definitions of velocity, with only $w^e_{50}$ recovering the observed VF.

Like the work presented in this paper, \citet{Maccio2016} also used cosmological zoomed simulations, the NIHAO suite, to make mock HI profiles, and showed that their measured $w^e_{50}$ could reproduce the observed VF.  \citet{Maccio2016} followed a similar analysis as in this paper, and both works use galaxies simulated with the code {\sc Gasoline}, but the simulations vary in terms of details.  A slightly lower resolution in most of the NIHAO galaxies prevents the use of H$_2$-based star formation as used here, but NIHAO includes a prescription for early stellar feedback (feedback from young massive stars that is deposited prior to the first SNII from any given star particle).  A detailed comparison of mock observed velocities at a given $v_{max,dmo}$ shows that the mock velocities in NIHAO are lower than in this work.  Perhaps because of this, \citet{Maccio2016} need not consider completeness in order to reproduce the observed VF; the lower velocities of $w^e_{50}$ alone are enough  to allow the NIHAO galaxies to match the data (and may even slightly over-reduce the velocities in the lowest halos; see their figure 3).

Thus, this work and both \citet{Brook2016} and \citet{Maccio2016} have concluded that the difference between $v_{max,dmo}$ and $w^e_{50}$ is the primary reason for the disagreement between theory and observations.  An apples-to-apples comparison between models and real galaxies alleviates the tension.

On the other hand, \citet{Trujillo2016} attempt to correct observed velocities to their underling $v_{max}$.  Using a sample of galaxies with resolved HI rotation curves from \citet{Papastergis2016}, they fit $v_{out}$ to both NFW and cored rotation curve models in order to infer the true $v_{max}$ of each galaxy.  This correction can then be applied to galaxies with unresolved HI velocities of similar baryonic mass.  However, they conclude that there is not enough of a shift to resolve the discrepancy between the theoretical and observed VF, even when the effects of dark matter core creation are taken into account.  

To reconcile the work of \citet{Trujillo2016} with the conclusions in this paper, \citet{Brook2016}, and \citet{Maccio2016}, the correction from observed $v_{out}$ to $v_{max}$ must fail.  Mock resolved HI rotation curves of the simulated galaxies should, in principal, be able to address this question.  However, results so far are inconclusive.  \citet{Read2016b} made mock HI rotation curves of two simulated dwarf galaxies and tested the conditions under which they could reliably recover the model halo masses.  They found that starburst and post-starburst dwarf galaxies have large HI bubbles that push the rotation curve out of equilibrium, and that galaxies viewed near face-on also presented problems, but could otherwise recover their model inputs (as long as they used a model with a dark matter core).  They concluded that a carefully selected sample should allow for a reliable recovery of true halo masses.  In \citet{Read2016c} they applied their method to 19 observed galaxies, and derived a stellar mass-to-halo mass relation in agreement with abundance matching results for field galaxies, concluding that there are no dwarf galaxy problems in CDM.  On the other hand, \citet{Verbeke2017} failed to recover the true $v_{circ}$ of any of their 10 dwarf galaxies (from the {\sc Moria} simulation suite) when producing mock HI rotation curves.  They conclude that the disks of dwarfs are simply too thick, combined with feedback causing significant structure and disequilibrium so that the HI rotation curve fails to be a good measure of the underlying gravitational potential.  Given the mixed results, more work in this area is required.

\subsection{Dark Matter Cores}

\begin{figure}[htbp]
\includegraphics[angle=0,width=\columnwidth]{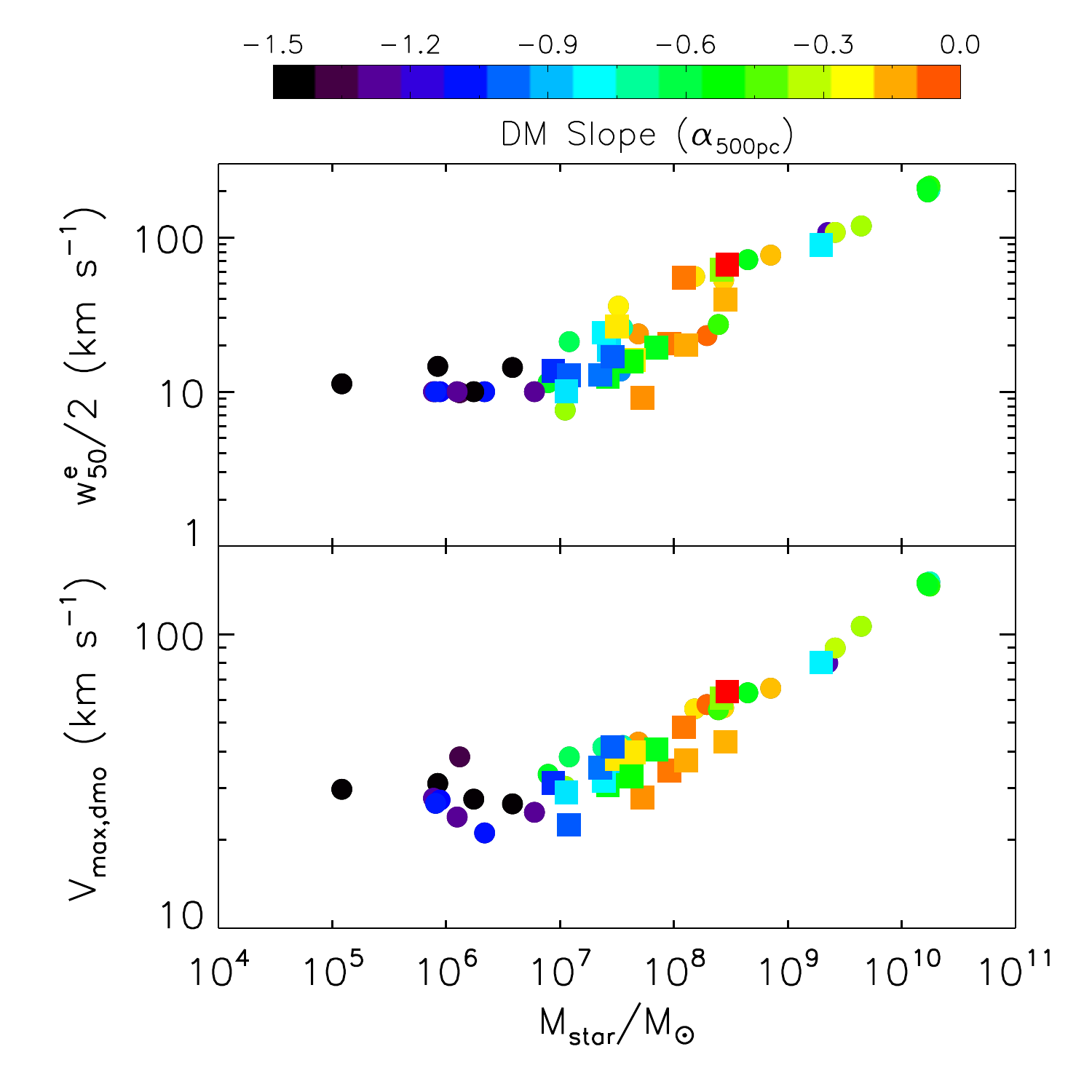}
\caption{Velocities derived from the simulations as a function of the stellar mass in the baryonic runs. As in previous figures, squares represent subhalos of larger galaxies while circles are central galaxies, and the data are color coded corresponding to the slope of the DM density profile in the baryonic version of the run.
{\it Top}: The mock ``observed'' velocity, \wfiftye/2, of the baryonic version of the galaxy plotted against its stellar mass.  For gas poor galaxies, \wfiftye is the stellar velocity dispersion, while $w^e_{50}$ is derived from the mock HI data cubes for gas-rich galaxies. 
{\it Bottom}: The $v_{max,dmo}$ of the DM-only version of the galaxy plotted against the stellar mass in the baryonic version.   
While the flattening of \wfiftye/2 at low galaxy masses has previously been attributed to DM core creation (see text), the fact that the trend remains in the bottom panel (that uses properties of the DM-only runs that retain a cuspy DM profile) indicates that the trend cannot be due to core creation. }
\label{trends}
\end{figure}

A recent analysis by \citet{Brook2015} also examined the effects of dark matter core creation on the observed galaxy VF, comparing to the Local Volume VF derived in \citet{Klypin2015}. The top panel of Figure~\ref{trends} shows the measured velocity dispersion (for HI poor galaxies) or edge-on $w^e_{50}$/2 (for HI rich galaxies) versus the stellar mass in the simulated galaxies.  The flattening of \wfiftye/2 below $\sim$10$^7$ M$_{\odot}$ is attributed to core creation in \citet{Brook2015}.  This flattening is not reproduced in their models with an NFW profile (they examine galaxies down to 10$^6$ M$_{\odot}$ in stellar mass). Only their model that includes dark matter core creation reproduces this flattening. 
Although we also find this flattening to occur at $\sim$10$^7$ M$_{\odot}$, the bottom panel of Figure~\ref{trends} demonstrates that this flattening in velocity cannot be due to core creation, as the trend is found in DM-only runs as well.

While \wfiftye/2 is a quantity derived from the baryonic simulations, the bottom panel of Figure~\ref{trends} shows $v_{max,dmo}$ plotted against the stellar mass of the galaxies in the baryonic version of the runs.  Recall that $v_{max,dmo}$ is a quantity derived from the DM-only versions of the galaxies.  DM core creation requires the presence of baryons, and hence cores cannot form in the DM-only runs.  The galaxies in the DM-only runs retain a steep, cuspy DM density profile.  Despite the steep inner profile, the flattening of the trend at low stellar masses persists in each panel.  Hence, core creation cannot be responsible for the flattening.  This is contrary to the conclusions in \citet{Brook2015}.  Reinforcing this conclusion, the data points in Figure~\ref{trends} are again color coded by the slope of the DM density profile in the baryonic version of the runs.  While cored galaxies tend to cluster in a given stellar mass range, they do not appear to play a role in the flattening of the trend below stellar masses of 10$^7$ M$_{\odot}$.

We note that the flattening below stellar masses of 10$^7$ M$_{\odot}$ is consistent with observational data \citep[see figure 1 of][]{Klypin2015}, which show a roughly constant velocity of $\sim$10 km s$^{-1}$ and tend to be in dispersion supported galaxies.  As previously discussed in Section 3.2, our faintest simulated galaxies have velocity dispersions $\sim$10 km s$^{-1}$, consistent with the observations.  This is a more direct comparison to the observations than presented in \citet{Brook2015}, where they measured $v_{circ}$ of a model galaxy at the radius that best reproduced $w_{50}$ values.

We offer a different interpretation for the flattening of velocities at low galaxy masses: the steep relation between M$_{star}$ and $v_{max,dmo}$ (or, equivalently, between M$_{star}$ and $M_{halo}$) at low halo masses.  As has been noted by previous authors \citep[e.g.,][]{Ferrero2012}, the steep relation at low halo masses suggests that galaxies over a wide range of stellar masses (10$^6$-10$^8$ M$_{\odot}$) reside in nearly the same host halo mass \citep[$\sim$10$^{10}$ M$_{\odot}$, though see][]{Read2016c}.  The bottom panel of Figure~\ref{trends} confirms that this trend also occurs in our simulations.  All of the low stellar mass galaxies reside in a narrow range of $v_{max,dmo}$ (or equivalently, $M_{halo}$).  This will lead them to have similar observed velocities as well, as seen in the top panel.

We note that unlike \citet{Brook2015}, our results are not in conflict with the conclusions in \citet{Brook2014}. In that paper, the role of core creation on the Too Big to Fail Problem \citep{Boylan-kolchin2011, Garrison-Kimmel2014} was explored using an analytic model (not simulations) for galaxies that had halo masses determined using stellar velocity dispersions at their half light radii.  In all cases, the half light radius is $\lesssim$1 kpc.  As we have seen, core creation {\it can} reduce the velocities of dwarfs interior to 1 kpc, and will thus alter the derived masses (though the magnitude of the reduction may not be as significant as \citet{Brook2014}  predicted for dwarf Irregulars, since they neglected gas in the inner kpc).  HI, on the other hand, can extend to much larger radii than typical half light radii, and eliminates any impact of dark matter cores on the measured velocity \citep[see also][]{Papastergis2017}.

\section{Conclusions}

In this work, we have used high resolution cosmological simulations of individual galaxies in order to resolve the discrepancy between the observed galaxy VF and the predicted VF within $\Lambda$CDM. In particular we study the apparent dearth of observed low velocity galaxies. 

To ensure that the simulated galaxies have realistic sizes and gas contents, and thus can be used to interpret observations, we verified that the simulated galaxies with baryons match observed scaling relations.  In particular, the simulations match the HI sizes of galaxies as a function of velocity and the baryonic Tully-Fisher relation. 

We use these realistic galaxies to generate mock ``observed'' velocities.  For galaxies with M$_{HI} > 10^6$ M$_{\odot}$, we produce mock HI datacubes and derive a characteristic velocity using the width of the HI profile at 50\% of the peak height, \wfifty.  This is the velocity commonly used to generate the observed galaxy VF \citep{Zwaan2010, Papastergis2011, Klypin2015}. For gas poor galaxies, we follow the procedure of \citet{Klypin2015} and use stellar velocity dispersion. When the ``observed'' velocities from baryonic simulations are compared to theoretical velocities (derived from the maximum circular velocity of matched counterpart halos in dark matter-only simulations), we find that there is a systematic shift in dwarf galaxies to lower velocities (see Figure \ref{fig4}).  The magnitude of this velocity shift, combined with a proper accounting of luminous halos, reconciles the observed VF with the theoretical VF (Figure \ref{fig5}).

Thus, there are two primary considerations necessary to bring the theoretical VF into agreement with the observed VF.  First, to match the observed VF at velocities below \wfifty ~$\sim$40 \kms, the fraction of luminous halos must be accounted for. If a halo does not host a luminous galaxy, it will remain undetected in current surveys, lowering the observed number of galaxies at low velocities compared to theoretical expectations that allow all halos to host a detectable galaxy. Here, we calculated the luminous fraction for halos with M$_{*} > 10^6$ M$_{\odot}$, which corresponds to the lower luminosity limit used to calculate the observed VF in \citet{Klypin2015}.  The fraction of luminous halos drops precipitously below 40 \kms.  Without considering this effect, the velocity difference alone between our mock observations and theoretical velocities is not sufficient to reproduce the observed VF at the low velocity end.  We note that previous work on this subject did not explicitly consider the fraction of luminous halos \citep[e.g.,][]{Brook2015, Maccio2016}.

Second, to match the observed VF it is necessary to derive a relationship between observed characteristic velocities of galaxies and theoretical velocities for halos.  We have demonstrated here that this relationship shifts the predicted VF into agreement with the current observations.  The source of the velocity shift in dwarf galaxies is a combination of factors:

\smallskip
 (1) The primary shift that makes observed velocities lower than theoretical velocities in dwarf galaxies is due to the fact that the velocity tracer (typically HI) does not trace the full potential wells of dwarfs.  That is, the outermost HI is still on the rising part of the rotation curve \citep{Catinella2006, deblok2008, Swaters2009, Oh2011}.  We demonstrate this in the top panel of Figure \ref{fig9}, where we explicitly found the circular velocity at the radius that the HI surface density dropped below 1 M$_{\odot}$/pc$^2$, i.e., the outermost observable rotation velocity, $v_{out}$, in simulated galaxies with baryons.  The top panel of Figure \ref{fig9} shows that $v_{out}$ underpredicts the maximum value of the circular velocity in dwarf galaxies.

\smallskip
 (2) For galaxies that derive a characteristic velocity using \wfifty, there is an additional reduction in observed velocity.  We demonstrate this in the second panel of Figure \ref{fig9}, where we compare the velocity results derived from \wfifty ~to $v_{out}$.  Although $v_{out}$ was already lower than the true maximum velocity of a galaxy's dark matter halo, \wfifty ~can be an additional 50\% lower than $v_{out}$.  This is because the HI line profile shape in dwarfs tends to be gaussian.  Measuring at a lower peak height, 20\%, instead agrees with $v_{out}$ (third panel of Figure \ref{fig9}).  To date, essentially all observed VF measurements have been made with \wfifty ~rather than $w_{20}$ because typical signal-to-noise ratios generally prevent a reliable measurement of $w_{20}$.

\smallskip
 (3) For galaxies with HI sizes under $\sim$3 kpc, an additional reduction in velocity can occur if the galaxy has a dark matter core.  Typical core sizes found in simulations are on the order of 1-2 kpc, and we demonstrate in Figure \ref{h516} that core creation reduces the overall circular velocity in the very center of cored galaxies.  However, if the characteristic rotational velocity is derived at a larger radius, then the measured circular velocity is usually comparable to the expected theoretical velocity.  We attempted to quantify the contribution of core creation to the reduction in velocity in Figure \ref{cores}.  There, we compared the circular velocity of halos in both the baryonic and contracted (dark matter-only + baryons) models at $R_{out}$.  This removes the contribution to the reduction in velocity due to being on the rising part of the rotation curve, and avoids the reduction due to \wfifty.  Figure \ref{cores} shows that galaxies with HI sizes $<$ 3 kpc typically have lower circular velocities than the contracted dark matter models, by up to 40\%.  This reduction is comparable to the reduction in velocity from measuring on the rising part of the rotation curve alone (see top panel of Figure~\ref{fig9}).  Hence, core creation leads to a further reduction in observed velocities for galaxies with R$_{out} < 3$ kpc.

\smallskip
Overall, we have demonstrated in this paper that we can start with the {\it abundance} of dwarf galaxies predicted in $\Lambda$CDM and reconcile the theoretical predictions with the observed VF.  We do this by properly accounting for the relation between characteristic velocities derived from observations and the characteristic velocities typically derived from theory, and by accounting for the fraction of observable halos detectable in current VF studies.  We conclude that there is no missing dwarf problem in $\Lambda$CDM.

\acknowledgments

AB acknowledges support from National Science Foundation (NSF) grant AST-1411399. EP is supported by a NOVA postdoctoral fellowship to the Kapteyn Astronomical Institute.  FG was partially supported by NSF grant AST-1410012, HST theory grant AR-14281, and NASA grant NNX15AB17G.
Resources supporting this work were provided by the NASA High-End Computing (HEC) Program through the NASA Advanced Supercomputing (NAS) Division at Ames Research Center.  This work was performed in part at the Aspen Center for Physics, which is supported by National Science Foundation grant PHY-1066293.

%\bibliographystyle{apj}
%\bibliography{master} 

\end{document}